\documentstyle[aps,prb,epsfig,psfig,floats,color]{revtex}

\begin{document}                                                             
%\baselineskip 0.2truecm                                                      
\newcommand{\beq}{\begin{equation}}
\newcommand{\eeq}{\end{equation}}
\newcommand{\bea}{\begin{eqnarray}}
\newcommand{\eea}{\end{eqnarray}}
\newcommand{\dfrac}{\displaystyle\frac}                                    
\newcommand{\disp}{\displaystyle}                                    
\newcommand{\mbf}{\mathbf}
\newcommand{\dint}{\displaystyle\int}
\renewcommand{\u}{\underline}                                        
\renewcommand{\o}{\overline}                                        
\newcommand{\Ptwo}{{\mathcal P}_2}
\newcommand{\bi}{\begin{itemize}}
\newcommand{\ei}{\end{itemize}}
\newcommand{\thint}{\widetilde{d\theta}}
\newcommand{\thprint}{\widetilde{d\theta'}}
\newcommand{\angint}{\int\! \thint\ }
\newcommand{\totint}{\int\! dl\ \thint\ }
\newcommand{\lint}{\int\! dl\ }
%{\dint \dfrac{1}{2}\ d\cos\theta\ }
\newcommand{\fexc}{\tilde{f}}
\newcommand{\fmom}{f_{\rm{ mom}}}
\newcommand{\fid}{f_{\rm{ id}}}
\newcommand{\rhoa}{\rho^{(a)}}
\newcommand{\va}{v^{(a)}}
\newcommand{\rhob}{\rho^{(b)}}
\newcommand{\vb}{v^{(b)}}
\newcommand{\parent}{\rho^{(0)}}
\newcommand{\normparent}{P^{(0)}}
\newcommand{\Ilm}{I_1}
\newcommand{\In}{J}
\newcommand{\tI}{I'}
\newcommand{\II}{{\tilde I}}
\newcommand{\I}{^{\rm{I}}}
\newcommand{\N}{^{\rm{N}}}
\newcommand{\phiparent}{\phi^{(0)}}
\newcommand{\bfig}{\begin{figure}[htb]\begin{center}}
\newcommand{\efig}{\end{center}\end{figure}}
\newcommand{\eqref}[1]{(\ref{#1})}
\newcommand{\eqq}[2]{~(\ref{#1},\ref{#2})}
\newcommand{\eqqq}[3]{~(\ref{#1},\ref{#2},\ref{#3})}
\newcommand{\eqm}[2]{(\ref{#1}-\ref{#2})}
\newcommand{\ie}{{\em i.e.}}
\newcommand{\eg}{{\em e.g.}}
\newcommand{\ft}{f(\theta)}
\newcommand{\order}{{\mathcal O}}

\newtheorem{thm}{Theorem}     [section]                                    
\newtheorem{definition}{Def.}   [section]                                    
\newtheorem{obs}{Remark}  [section]

\twocolumn[\hsize\textwidth\columnwidth\hsize\csname @twocolumnfalse\endcsname

\title{Isotropic-nematic phase equilibria of polydisperse hard rods:
The effect of fat tails in the length distribution}
\author{Alessandro Speranza, Peter Sollich}

\address{Department of Mathematics, King's College London,
Strand, London WC2R 2LS, U.K.
Email: peter.sollich@kcl.ac.uk}

\maketitle

\begin{abstract}
We study the phase behaviour of hard rods with length polydispersity,
treated within a simplified version of the Onsager model. We give a
detailed description of the unusual phase behaviour of the system when
the rod length distribution has a "fat" (e.g.\ log-normal) tail up to
some finite cutoff. The relatively large number of long rods in the
system strongly influences the phase behaviour: the isotropic cloud
curve, which defines the point where a nematic phase first occurs as density
is increased, exhibits a kink; at this point the properties of the
coexisting nematic shadow phase change discontinuously. A narrow
three-phase isotropic-nematic-nematic coexistence region exists near
the kink in the cloud curve, even though the length distribution is
unimodal. A theoretical derivation of the isotropic cloud curve and
nematic shadow curve, in the limit of large cutoff, is also given. The
two curves are shown to collapse onto each other in the limit. The
coexisting isotropic and nematic phases are essentially identical, the
only difference being that the nematic contains a larger number of the
longest rods; the longer rods are also the only ones that show any
significant nematic ordering. Numerical results for finite but large
cutoff support the theoretical predictions for the asymptotic scaling
of all quantities with the cutoff length.

\end{abstract}
\vspace*{0.5cm}
]

\section{Introduction}
Rod-like particles in suspension can undergo a phase transition
between an isotropic (I) phase without orientational order and a
nematic (N) phase where rods are preferentially oriented along the
so-called nematic axis. This transition has been observed
experimentally for chemical and biological
systems~\cite{Zocher25,BawPirBerFan36,BuiPatPhiLek93,%
BuiVelPatLek92,VanVanLek96}. Theoretically, the two main approaches
for analysing systems of rod-like particles are due to Maier and
Saupe~\cite{MaiSau58} and Onsager~\cite{Onsager49}. The Maier-Saupe
theory is based on a long range attractive interaction between
particles (originally conceived of as due to van der Waals forces) and
neglects density variations. Such a theory is therefore appropriate for
describing thermotropic liquid crystals, in which orientational phase
transitions are induced by changes in temperature.  The Onsager
theory, on the other hand, only considers hard core (short range)
interactions between particles. Temperature then
becomes just a trivial factor setting the energy scale, and the
Onsager theory therefore describes {\em lyotropics}: materials in
which the phase transition is driven by a change in density of the
system (at fixed temperature). The I-N phase transition is due, within
the Onsager theory, to a competition between the orientational
entropy---the tendency of rods to stay orientationally
disordered---and the packing entropy; the latter is due to the
excluded volume interaction, and higher for
aligned particles. The key simplification of Onsager theory is to
consider the ``Onsager limit'' of thin rods, where the ratio $D/L$ of
the diameter and the length of the rods tends to zero. The free energy
of the model can then be derived from a virial expansion truncated
after the first nontrivial term, as all higher order terms turn out to
be smaller by positive powers of $D/L$. To find the actual free energy
for a given rod number density, one needs to minimize over the
orientational distribution of the rods; once this is done, the I-N
phase transition can be obtained by a standard double-tangent
construction. Onsager solved the minimization problem by assuming a
parametric form for the orientational distribution function; even with
only one variational parameter, he obtained results for the densities
of the coexisting phases~\cite{Onsager49,LekCouVanDeb84} which are in
good agreement with the numerically exact solution obtained
later~\cite{KayRav78}.

A comparison between experimental systems and the theoretical
predictions of Onsager theory can be complicated due to \eg\ non-hard
interactions or particles that are not perfectly
rigid~\cite{BuiPatPhiLek93,BuiVelPatLek92,KhoSem81,KhoSem81_2}. A
further important factor is polydispersity, \ie\ a spread in particle
lengths and/or diameters~\cite{BuiLek93,VanVanLek96}, and we focus in
this paper on the effects that length polydispersity has on the phase
behaviour of hard rods. Polydispersity makes the analysis of the
problem more difficult, but also leads to much richer phase
behaviour~\cite{Sollich02}. Even in simple bi- and
tri-disperse systems, \ie\ mixtures of rods of two or three different
lengths, phase separation into two nematic (N-N) phases and
three-phase I-N-N coexistence have been predicted
theoretically~\cite{BuiLek93,LekCouVanDeb84,OdiLek85,VanMul96,VroLek93},
and observed experimentally~\cite{BuiLek93}. More generally, a
pronounced broadening of the coexistence region and fractionation of
the longer rods into the nematic phases are typical effects of length
polydispersity~\cite{BuiLek93,Chen94,Evans99,Sluckin89,VanVanLek96}.
These effects are also seen in simplified models, such as the
polydisperse Zwanzig model~\cite{ClaCueSeaSolSpe00,Zwanzig63}---where
rods are only allowed to point along one of three orthogonal
directions---or the lattice model developed by Flory et
al.~\cite{Flory56,FloAbe78,AbeFlo78}. Within the context of the
lattice model, exponential and Poisson~\cite{FloFro78,FroFlo78} as
well as Gaussian~\cite{MosWil82} length distributions were analysed,
with fractionation and a broadening of the coexistence region observed
in all cases. The bidisperse case was also studied~\cite{AbeFlo78}
and, despite the rather different theoretical approach, showed phase
behaviour qualitatively similar to that predicted within bidisperse
Onsager theory~\cite{BirKolPry86,BirKolPry88}.

It is clear from the above discussion
that more work is needed to understand the
effects of length polydispersity on the phase behaviour of hard rods,
especially with regard to the occurrence of the more ``exotic'' N-N
and I-N-N phase coexistences. We recently began to investigate these
questions, using the ``$\Ptwo$ Onsager model''~\cite{SpeSol_p2}. This
is obtained from the conventional Onsager theory by simplifying the
angular dependence of the excluded volume term in the free energy,
truncating a series expansion in Legendre polynomials after the first
nontrivial term, which contains the second order Legendre polynomial
$\Ptwo(\cos\theta)$. The simplicity of the resulting model enabled us
to investigate in detail the phase behaviour resulting from different
choices of rod length distributions. For a Schulz distribution---which
is unimodal, \ie\ has a single peak---fractionation and broadening of
the coexistence region were found as expected, but neither N-N nor
I-N-N coexistence appeared.  These more complex features did occur,
however, for bidisperse and bimodal distributions; the latter were
modelled by a mixture of two Schulz distributions peaked at different
lengths. For the bidisperse case the appearance of these features is
in encouraging qualitative agreement with the results of Onsager
theory, suggesting that the approximation made in constructing the
$\Ptwo$ model may not be crucial. A detailed analysis of the phase
diagram of the $\Ptwo$ Onsager model showed that N-N and I-N-N regions
only appeared if the ratio of the rod lengths was sufficiently large
while the fraction of long rods in the system remained small. (For the
bimodal case the individual peaks in the distribution also had to
remain relatively narrow.) This suggested to us that N-N and I-N-N
coexistence could also occur in unimodal length distributions that
contain a larger number of long rods than the Schulz distribution with
its exponentially decaying tail. We therefore investigate, in the
present paper, the behaviour of the $\Ptwo$ Onsager model for rod
length distributions with fat tails, \ie\ decaying less than
exponentially for large rod lengths. Our primary example will be the
log-normal distribution; like the Schulz distribution, this has most
of its weight around the average rod length, but contains a comparatively
larger number of much longer rods; it has also yielded interesting
results in previous work on length-polydisperse
homopolymers~\cite{Solc70,Solc75}. Our main result is that I-N-N phase
coexistence does indeed occur, although the topology of the phase
diagram is rather different from the bidisperse case: the I-N-N region
is very narrow in density and remains fully within the
isotropic-nematic region, without being bordered by an N-N coexistence
region.

The present paper is structured as follows. In Sec.~\ref{sec:model}
we give a brief description of the polydisperse $\Ptwo$ Onsager model
and of the phase coexistence equations. We also motivate why the
presence of fat tails should cause unusual phase behaviour. To make
the phase coexistence problem well-defined, a cutoff on the rod length
distribution needs to be introduced; this is of course also reasonable
physically.
Sec.~\ref{sec:numerical_res} contains our numerical results for the
phase diagram at finite cutoff, and we describe carefully the unusual
effects on phase behaviour caused by the presence of the long rods; it
turns out that such effects are largely confined to a region of low
densities where phase ordering would not yet be observed in systems
with ``conventional'' length distributions. The second major part of
the paper contains a theoretical treatment of the onset of nematic
ordering, in the limit of large length cutoff
(Sec.~\ref{sec:limit}). This reveals many unusual features; in the
limit of infinite cutoff, for example, both the densities and the rod
volume fractions of the isotropic cloud phase and the coexisting
nematic shadow phase coincide. The theoretical predictions are
compared to the results of our numerical calculations, and show good
agreement. We conclude in Sec.~\ref{sec:conclusion} with a summary and
outlook towards future work. Technical material is relegated to two
appendices.

\section{The polydisperse $\Ptwo$ Onsager model}\label{sec:model}

The $\Ptwo$ Onsager model is an approximate version of the 
Onsager model of hard rods. As in the Onsager theory, particles are
modelled as rigid spherocylinders with hard core
interaction. Phase transitions are driven by density rather than
temperature; the latter simply fixes the energy scale and can be set
to unity~\cite{Onsager49,Zwanzig63,ClaCueSeaSolSpe00,SpeSol_p2}.

We allow for length polydispersity, with the rod lengths $L$
distributed according to a length distribution $P(L)$, while assuming
that all rod diameters $D$ are equal. To be able to take the Onsager
limit of thin rods simultaneously for all rod lengths in $P(L)$, we
introduce a reference length $L_0$ and consider the limit $D/L_0\to 0$
while keeping the normalized lengths $l=L/L_0$ constant. 
The thermodynamic state of the system is specified by
the density distribution $\rho(l,\theta)$,
where $\rho(l,\theta)\,dl\,d\Omega/(4\pi)$ gives the number density of rods
with (normalized) lengths in the range $l\ldots l+dl$ and orientations
in a solid angle $d\Omega$ around any direction at an angle $\theta$
with the nematic axis~\cite{SpeSol_p2,ClaCueSeaSolSpe00}
We can then decompose $\rho(l,\theta)$ into
\begin{equation}\label{eq:decomposition_rho_theory}
\rho(l,\theta)=\rho(l)P_l(\theta)=\rho_0 P(l) P_l(\theta)
\end{equation}
where we have isolated the overall density $\rho_0$ and the normalized
length distribution $P(l)$. The angular probability
distributions $P_l(\theta)$ are normalized in such a way that
\begin{equation}\label{eq:normalization_ang_P_theory}
\angint P_l(\theta)=1
\end{equation}
using the shorthand
\[
\thint=\frac{1}{2}d\cos\theta
\]
and the convention that all angular integrations are over the range
$0\ldots \pi$. Conversely, the density $\rho_0$ is obtained by
integrating $\rho(l,\theta)$ over all $l$ and $\theta$,
\[
\rho_0=\totint\rho(l,\theta)=\lint\rho(l)
\]
The excess free energy of the model arises from the excluded volume
interaction of the rods. In the Onsager limit, the virial expansion
can be truncated after the second-order term. The simplification of
the $\Ptwo$ Onsager model consists in simplifying the angular
dependence of the resulting excluded volume term, by expanding in
Legendre polynomials~\cite{KayRav78} and truncating after the first
nontrivial term. If all densities are measured in units of
$1/[(\pi/4)DL_0^2]$, the excess free energy density then becomes
simply~\cite{SpeSol_p2}
\begin{equation}\label{eq:excess_Ptwo_theory}
\fexc=\frac{c_1}{2}\rho_1^2-\frac{c_2}{2}\rho_2^2
\end{equation}
where $c_1=2$ and $c_2=5/4$. We have also defined the two moments of
the density distribution
\begin{equation}\label{eq:rho1_theory}
\rho_1=\totint l\rho(l,\theta) =
\int dl\ l\rho(l)\angint P_l(\theta)
\end{equation}
and
\begin{equation}\label{eq:rho2_theory}
\rho_2=\totint l\Ptwo(\cos\theta) =
\int dl\ l\rho(l)\angint\Ptwo(\cos\theta)\ P_l(\theta)
\end{equation}
The first of these is just the rescaled rod volume fraction, $\rho_1 =
(L_0/D)\phi$, while $\rho_2$ contains information on the orientational
order in the system.  It is clear from Eq.~\eqref{eq:rho2_theory} that
for an isotropic phase ($P_l(\theta)\equiv 1$) $\rho_2$ vanishes,
while $0<\rho_2\leq\rho_1$ for a nematic phase, whose angular
distribution will be peaked around $\theta=0$ and $\theta=\pi$;
$\rho_2$ is therefore a natural orientational order parameter for the
system. In fact one can rewrite $\rho_2$ as
\beq\label{eq:rho2_S(l)}
\rho_2=\totint l\Ptwo(\cos\theta)\rho(l) P_l(\theta)=\int dl\ l\rho(l) S(l)
\eeq
if one defines the order parameter $S(l)$ as the average of the
second Legendre polynomial for rods of given length,
\beq\label{eq:S(l)}
S(l)=\angint \Ptwo(\cos\theta) P_l(\theta)
\eeq
Eq.~\eqref{eq:rho2_S(l)} can also be read as $\rho_2=\rho_0\langle
lS(l)\rangle$, where the average is taken over the normalized length
distribution $P(l)=\rho(l)/\rho_0$.

As usual the ideal part of the free energy, comprising the ideal
gas term and the entropy of mixing, is
\begin{equation}\label{eq:fid_theory}
\fid=\angint dl\ \rho(l, \theta)\left[\ln\rho(l, \theta)-1\right]
\end{equation}
Using the decomposition~\eqref{eq:decomposition_rho_theory} and the
normalization~\eqref{eq:normalization_ang_P_theory} we can then
write the total free energy as
\begin{equation}\label{eq:free_en_theory}
f=\dint dl\ \rho(l)\left[\ln\rho(l)-1\right]+\dint dl\ \rho(l)\angint
P_l(\theta)\ln P_l(\theta)+\fexc%\frac{c_1}{2}\rho_1^2-\frac{c_2}{2}\rho_2^2
\end{equation} 
with $\fexc$ given by Eq.~\eqref{eq:excess_Ptwo_theory}.

\subsection{Phase coexistence equations}\label{sec:phase_eq}

Since the rod orientations -- as opposed to their lengths -- are not
conserved, the orientational probability distributions $P_l(\theta)$
for a given density distribution $\rho(l)$ over rod lengths have to be
found by minimizing the free energy~\eqref{eq:free_en_theory}, subject
to the constraints~\eqref{eq:normalization_ang_P_theory}.
Introducing appropriate Lagrange
multipliers $\kappa(l)$ we obtain the condition
\begin{eqnarray*}\label{eq:minimization_theory}
\frac{\delta}{\delta P_l(\theta)}\left(f+\int dl\ \kappa(l)\angint
P_l(\theta)\right)=\\
\rho(l)\left[\ln P_l(\theta)+1\right]+l\rho(l)\left[c_1\rho_1-c_2\rho_2\Ptwo\right]+\kappa(l)=0
\end{eqnarray*}
Solving for $P_l(\theta)$ gives
\begin{equation}\label{eq:P_l_theory}
P_l(\theta)=\frac{\exp(lc_2\rho_2\Ptwo)}{\angint
\exp(lc_2\rho_2\Ptwo)}
\end{equation}
where we have introduced the shorthand $\Ptwo=\Ptwo(\cos\theta)$.
This, together with~\eqref{eq:rho2_theory} gives a self-consistency
equation
\beq\label{eq:self_cons}
\rho_2=\int dl\ l\rho(l)\frac{\angint\Ptwo\exp(lc_2\rho_2\Ptwo)}
{\angint\exp(lc_2\rho_2\Ptwo)}
\eeq
which can be solved for $\rho_2$. 

To calculate phase coexistences, we need the expression for the
chemical potentials and for the osmotic pressure. The chemical
potentials are obtained by taking a derivative of the free energy with
respect to the density distribution $\rho(l)$,
\begin{eqnarray}\nonumber
\mu(l)&=&\frac{\delta f}{\delta\rho(l)}\\
&=&\ln\rho(l)-\ln\angint\exp(-lc_1\rho_1+lc_2\rho_2\Ptwo)\label{eq:mu_theory}
\end{eqnarray}
where we do not need to differentiate explicitly with respect to
the $P_l(\theta)$ since they are chosen to minimize the free energy.
The osmotic pressure, derived from the Gibbs-Duhem relation, is
\begin{equation}\label{eq:Pi_theory}
\Pi=-f+\int dl\ \rho(l)\mu(l)=\rho_0+\frac{c_1}{2}\rho_1^2-\frac{c_2}{2}\rho_2^2
\end{equation}
Equality of the chemical potentials in a set of coexisting phases
labelled by $a=1\ldots P$ then leads to the following expression for the
length distribution
\begin{equation}\label{eq:phase_distrib_theory1}
\rhoa(l)=R(l)\angint\exp(-lc_1\rhoa_1+lc_2\rhoa_2\Ptwo)
\end{equation}
where $R(l)$ is a function common to all phases. It can be obtained
by imposing the lever rule or particle number conservation,
\begin{equation}\label{eq:lever_rule_theory}
\sum_a\va\rhoa(l)=\parent(l)
\end{equation}
where $\parent(l)$ is the overall or `parent' density distribution of
the system and $\va$ is the fraction of the system volume
occupied by phase $a$. This gives
\begin{equation}\label{eq:R(l)_theory}
R(l)=\frac{\parent(l)}{\sum_a\va\angint\exp(-lc_1\rhoa_1+lc_2\rhoa_2\Ptwo)}
\end{equation}
Using Eqs.\eqq{eq:P_l_theory}{eq:phase_distrib_theory1}, the full
density distributions over lengths and orientations are therefore
\beq\label{eq:phase_distrib_theory}
\rhoa(l,\theta)=\frac{\parent(l)\exp(-lc_1\rhoa_1+lc_2\rhoa_2\Ptwo)}
{\sum_b v^{(b)}\int\thprint\ \exp(-lc_1\rhob_1+lc_2\rhob_2\Ptwo')}
\eeq
We thus have a closed system of equations whose solutions determine the
phase behaviour of the $\Ptwo$ Onsager model.  For each of the $P$
phases we have 3 unknowns, $\rhoa_1$, $\rhoa_2$ and $\va$, giving $3P$
unknowns in total.  For each phase, $\rhoa_1$ and $\rhoa_2$ obey the
equations obtained by substituting Eq.~\eqref{eq:phase_distrib_theory}
into Eqs.\eqq{eq:rho1_theory}{eq:rho2_theory}; for $\rhoa_2$ this
leads back to Eq.~\eqref{eq:self_cons}.  The remaining $P$
equations are given by the equality of the osmotic pressure in all
phases ($P-1$ equations) and the normalization condition $\sum_a
v^{(a)}=1$. The lever rule and the equality of chemical potentials are
already ensured by having density distributions in the different
phases of the form~\eqref{eq:phase_distrib_theory}.

So far everything holds for arbitrary parent distributions
$\parent(l)$. We will denote by $\parent_0$ the overall parent number
density, so that $\normparent(l)=\parent(l)/\parent_0$ is the normalized
parent rod length distribution. We always assume that the average rod
length in the parent is one; a different value could be absorbed into
the reference length $L_0$.

\subsection{Fat-tailed distributions}\label{sec:log-normal}

We now motivate why the phase behaviour of systems with rod length
distributions with fat (less than exponentially decaying) tails should
be unusual. To do so, it is useful to focus on the onset of the
isotropic-nematic coexistence (the so called ``isotropic cloud
point'').  Equality of the chemical potentials~\eqref{eq:mu_theory}
between an isotropic ($\rho\I(l)$) and a nematic ($\rho\N (l)$) phase
gives
\beq\label{eq:nematic_distr}
\rho\N (l)=\rho\I(l)e^{\beta
l}\angint\exp\left[c_2\rho_2\N l(\Ptwo-1)\right]
%=\rho\I(l)e^{\beta l}\angint e^{lf(\theta)} 
\end{equation}
where we have defined
\beq\label{eq:beta_def}
\beta=-c_1\left(\rho_1\N -\rho_1\I\right)+c_2\rho_2\N 
\eeq
and used $\rho_2\I=0$. At the isotropic cloud point, \ie\ at the onset
of the phase coexistence between the isotropic parent and an
infinitesimal amount of nematic ``shadow'' phase,
$\rho\I(l)=\parent(l)$. Moreover, $\beta$ should be positive since the
nematic phase will have the larger rod volume fraction, $\rho_1\N
>\rho_1\I$. The angular integral in Eq.~\eqref{eq:nematic_distr},
finally, is bounded by unity since $\Ptwo-1\leq 0$, and a Gaussian
approximation around $\theta=0$ that is valid for large $l$ shows that
it varies only as a power law with $l$. Taken together, these facts imply
that as soon as the parent distribution $\parent(l)$ is less than
exponentially decaying with $l$, the nematic length distribution would
diverge for large $l$. In order to ensure convergent integrals for the
density $\rho\N _0$ and the rod volume fraction $\rho\N _1$ we will
therefore need to impose a maximum length cutoff $l_m$ on the parent
distribution. The need for such a cutoff suggests---and we will find
this confirmed below---that even though there are only a very small
number of long rods (since the integral $\lint\parent(l)$ converges)
they can dominate the onset of phase coexistence from the isotropic side.
Notice that, coming from the nematic, no such effects are expected.
At the nematic cloud point, $\rho\N(l)=\parent(l)$. Inverting
Eq.~\eqref{eq:nematic_distr} one sees then that the length
distribution in the isotropic shadow phase is well behaved even for
$l_m\to\infty$, as $\beta$ is still positive and the angular integral
is harmless.

Motivated by these insights, we consider fat-tailed parent
distributions in the rest of this paper. We expect, and indeed find,
that the presence of the fat tail, and the value of the cutoff $l_m$,
will have a strong influence on the isotropic cloud point and the
corresponding nematic shadow, while leaving the nematic cloud point
essentially unaffected. Although many of our results are valid for
general fat-tailed distributions, we will normally focus on log-normal
distributions for definiteness.

%%%%%%%%%%%%%%%%%%%%%%%%%%PHASE DIAGRAM%%%%%%%%%%%%%%%%%%%%%%%%%%%%%%%%%%%%%%%

\section{Numerical results}\label{sec:numerical_res}

In this section we describe the unusual phase
behaviour of systems with a
log-normal parent length distribution
\beq\label{eq:log-normal}
\normparent(l)=\frac{1}{\sqrt{2\pi w^2}l}\exp\left[-\frac{(\ln
l-\mu)^2}{2 w^2}\right]
\eeq
with a finite cutoff, $l\leq l_m$. The parameters $\mu$ and $w$ are
chosen such that the average rod length is $\langle l\rangle = 1$ and
the polydispersity $\sigma$, \ie\ the normalized standard deviation of
distribution of rod lengths defined through
\beq
\label{eq:sigma}
\sigma^2=\frac{\langle l^2\rangle}{\langle l\rangle^2}-1
\eeq
has the desired value; explicitly, this gives $w^2=\ln(1+\sigma^2)$
and $\mu=-w^2/2$. Note that for finite cutoff $l_m$, $\normparent(l)$
as given will not be normalized precisely to one, and $\langle l
\rangle$ and $\langle l^2\rangle$ will differ slightly from the
desired values of 1 and $1+\sigma^2$. However, even for modest cutoffs
the deviations are very small. For instance, at cutoff $l_m=50$ and
$\sigma=0.5$, the integrals $\lint \normparent(l)l^n$ differ from
their $l_m\to\infty$ values $1$, $1$ and $1+\sigma^2$ by values of
order $10^{-17}$, $10^{-15}$, and $10^{-13}$. In the following we will
neglect these small corrections.

Due to the numerical difficulties introduced by polydispersity, and
compounded by the presence of a fat tail in the parent distribution,
we used the {moment method}~\cite{SolCat98,SolWarCat01,Warren98} to
solve numerically an approximate system of phase coexistence
equations, applying the recently developed {adaptive
method}~\cite{ClaCueSeaSolSpe00,SpeSol_p2} to keep deviations from
the exact solution as small as possible. We found that controlling
such deviations was rather more difficult than for more
``well-behaved'' rod length distributions, and therefore decided to
integrate the moment method with the exact solution. In this mixed
method, we effectively use the adaptive moment method to produce a
good starting point for the numerical solution of the exact phase
coexistence conditions. This method allows us to calculate the exact
phase behaviour, while avoiding the convergence problems that result
from a direct numerical attack on the exact phase coexistence
conditions.

\subsection{Overview of the limiting theory}\label{sec:theory_sum}

Before showing the numerical results at finite cutoff $l_m$, it is
useful to anticipate some of the analytical results which we will
derive later on, in Sec.~\ref{sec:limit}, for the large
cutoff limit. This helps to understand and physically interpret some
of the unusual features of the phase diagram obtained at finite
cutoff.

The main feature affecting the physics of the system is obviously the
presence of a large (relative to more strongly decaying distributions,
\eg\ of Schulz type) number of long rods. Because of the
weighting with $l$ in the moment
densities\eqq{eq:rho1_theory}{eq:rho2_theory} appearing in the excess
free energy, it is plausible that these long rods can drive the phase
ordering. Indeed, our theory will show that rods with $l$ of the order
of the cutoff $l_m$ become strongly ordered already when the nematic
phase first appears, and that this ordering drives the onset of phase
coexistence to lower densities. The isotropic cloud point therefore
depends on the cutoff, moving to lower densities as $l_m$ increases.
However, the ordering of the longest rods is not sufficient to
precipitate general order in the nematic phase. In fact, we find that
at the relatively low isotropic cloud point densities caused by the
presence of the long rods, the short rods with $l$ of order unity are
not yet able to order, staying completely disordered in the
limit $l_m\to\infty$. The system thus has an unusual initial phase
separation, in which the nematic phase is distinguishable from the
isotropic phase only because it contains a somewhat larger fraction of
the longest rods, which are also nematically ordered; for the short
rods, the density distribution is essentially identical in both
phases. At finite $l_m$ the contribution of the longest rods causes
the nematic phase to have a larger rod volume fraction and average rod
length than the isotropic phase. As $l_m$ increases, however, the
differences turn 
out to vanish, and the number density and rod volume fraction of the
nematic shadow phase become identical to those of the isotropic cloud
phase. In a plot of these quantities
versus the polydispersity $\sigma$, the resulting cloud and shadow
curves therefore coincide for $l_m\to\infty$. In fact, we will obtain
this limiting curve explicitly as
\[
\rho_0=\rho_1=\frac{1}{4}\frac{1}{(\sigma^2+1)}
\]
Our theory also predicts the leading terms in the approach to this
asymptotic behaviour as $l_m$ increases. The number density and volume
fraction of the isotropic and the nematic phase turn out to differ, in
the limit, by terms proportional to $\rho_2\N$; here $\rho_2\N$ is the
value of the moment density~\eqref{eq:rho2_theory} in the nematic
phase, which for a log-normal parent scales as $(\ln
l_m)/l_m^{1/2}$. The parameter $\beta$ in Eq.~\eqref{eq:nematic_distr}
turns out to scale like $\left(\rho_2\N \right)^2$ to leading
order. This decrease of $\beta$ to zero allows the theory to remain
well-defined, with finite values for the number density and rod volume
fraction of the nematic for $l_m\to\infty$. In fact, the scaling of
$\beta$ is such that the two factors $\exp(\beta l)$ and
$\rho\I(l)=\parent(l)$ in Eq.~\eqref{eq:nematic_distr} approximately
cancel for $l=l_m$, with their product varying only as a power law in
$l_m$. In our numerical results at finite $l_m$, this feature will be
visible as a peak at $l=l_m$ in the density distribution of the
nematic shadow phase. The power law scaling of the weight of the peak
is such that its contribution to the moment densities $\rho\N_0$ and
$\rho\N_1$ becomes negligible for $l_m\to\infty$, as pointed out
above. On the other hand, higher moments of the nematic density
distribution, $\int dl\, \rho\N(l)l^n$ with $n\geq 2$, are dominated by
the contribution from the peak and actually diverge with
$l_m$. Similar unusual features also appear in, for example, the
Flory-Huggins theory of polymers with log-normal chain length
distribution~\cite{Solc70,Solc75}.

\subsection{Overall phase diagram topology}\label{sec:phase_diagram}

In Fig.~\ref{fig:phased_log_new} we show the phase diagram
obtained for a log-normal parent distribution with cutoff $l_m=100$. Plotted
is the density of the parent at which phase transitions occur; the
lines separating one- and two-phase regions are therefore cloud
curves.

\bfig
\begin{picture}(0,0)%
\includegraphics{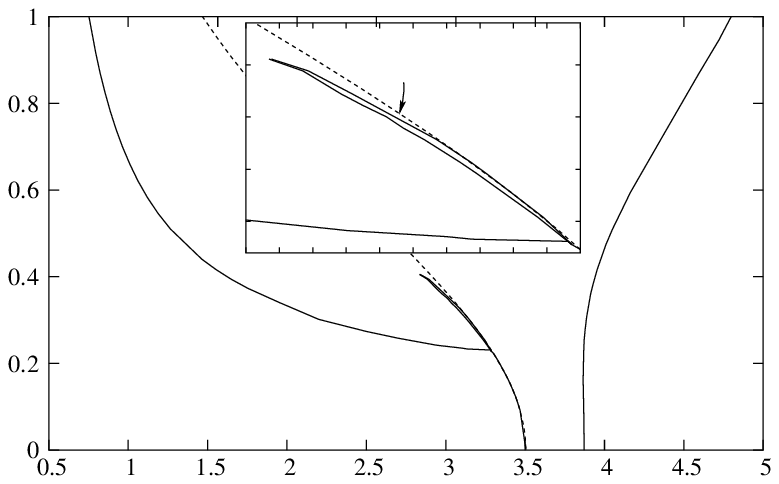}%
\end{picture}%
\setlength{\unitlength}{2565sp}%
\begingroup\makeatletter\ifx\SetFigFont\undefined
% extract first six characters in \fmtname
\def\x#1#2#3#4#5#6#7\relax{\def\x{#1#2#3#4#5#6}}%
\expandafter\x\fmtname xxxxxx\relax \def\y{splain}%
\ifx\x\y   % LaTeX or SliTeX?
\gdef\SetFigFont#1#2#3{%
  \ifnum #1<17\tiny\else \ifnum #1<20\small\else
  \ifnum #1<24\normalsize\else \ifnum #1<29\large\else
  \ifnum #1<34\Large\else \ifnum #1<41\LARGE\else
     \huge\fi\fi\fi\fi\fi\fi
  \csname #3\endcsname}%
\else
\gdef\SetFigFont#1#2#3{\begingroup
  \count@#1\relax \ifnum 25<\count@\count@25\fi
  \def\x{\endgroup\@setsize\SetFigFont{#2pt}}%
  \expandafter\x
    \csname \romannumeral\the\count@ pt\expandafter\endcsname
    \csname @\romannumeral\the\count@ pt\endcsname
  \csname #3\endcsname}%
\fi
\fi\endgroup
\begin{picture}(6076,3714)(976,-4161)
\put(976,-2086){\makebox(0,0)[lb]{\smash{\SetFigFont{8}{9.6}{rm}$\sigma$}}}
\put(4276,-4111){\makebox(0,0)[lb]{\smash{\SetFigFont{8}{9.6}{rm}{\color[rgb]{0,0,0}$\parent_0$}%
}}}
\put(6376,-2686){\makebox(0,0)[lb]{\smash{\SetFigFont{8}{9.6}{rm}{\color[rgb]{0,0,0}N}%
}}}
\put(2401,-2911){\makebox(0,0)[lb]{\smash{\SetFigFont{8}{9.6}{rm}{\color[rgb]{0,0,0}I}%
}}}
\put(3826,-2611){\makebox(0,0)[lb]{\smash{\SetFigFont{8}{9.6}{rm}{\color[rgb]{0,0,0}I+N$_1$}%
}}}
\put(5026,-2686){\makebox(0,0)[lb]{\smash{\SetFigFont{8}{9.6}{rm}{\color[rgb]{0,0,0}I+N$_2$}%
}}}
\put(4929,-1193){\makebox(0,0)[lb]{\smash{\SetFigFont{8}{9.6}{rm}{\color[rgb]{0,0,0}I+N$_2$}%
}}}
\put(4123,-955){\makebox(0,0)[lb]{\smash{\SetFigFont{8}{9.6}{rm}{\color[rgb]{0,0,0}I+N$_1$+N$_2$}%
}}}
\put(3596,-1750){\makebox(0,0)[lb]{\smash{\SetFigFont{8}{9.6}{rm}{\color[rgb]{0,0,0}I+N$_1$}%
}}}
\put(3132,-1671){\makebox(0,0)[rb]{\smash{\SetFigFont{8}{9.6}{rm}{\color[rgb]{0,0,0}0.3}%
}}}
\put(3132,-899){\makebox(0,0)[rb]{\smash{\SetFigFont{8}{9.6}{rm}{\color[rgb]{0,0,0}0.4}%
}}}
\put(3662,-2411){\makebox(0,0)[b]{\smash{\SetFigFont{8}{9.6}{rm}{\color[rgb]{0,0,0}2.9}%
}}}
\put(4156,-2411){\makebox(0,0)[b]{\smash{\SetFigFont{8}{9.6}{rm}{\color[rgb]{0,0,0}3}%
}}}
\put(4650,-2411){\makebox(0,0)[b]{\smash{\SetFigFont{8}{9.6}{rm}{\color[rgb]{0,0,0}3.1}%
}}}
\put(5143,-2411){\makebox(0,0)[b]{\smash{\SetFigFont{8}{9.6}{rm}{\color[rgb]{0,0,0}3.2}%
}}}
\end{picture}
\vspace*{0.3cm}
\caption{Phase diagram for a log-normal parent with cutoff
$l_m=100$. Shown is the polydispersity $\sigma$ plotted against the
density of the parent $\parent_0$ at which phase transitions
occur. From left to right we find a single-phase isotropic phase, the
I-N coexistence region and then the single-phase nematic; the
coexistence region is bounded from left and right by the isotropic and
nematic cloud curves, respectively.  Inside the I-N coexistence region
we observe a narrow three-phase I-N-N region. This is shown in more detail in
the inset and lies close to the isotropic cloud curve (dashed line) that
would result for a Schulz distribution. Above the three-phase region
one has, instead of N-N phase separation, a crossover
from a cutoff-dependent regime (located 
to the right of the isotropic cloud curve) to a regime in which the phase
behaviour is almost unaffected by $l_m$; the location of the crossover
is close to the dashed line, \ie\ the ``conventional'' (Schulz) cloud 
curve.}
\label{fig:phased_log_new}
\efig
Vertically, the phase diagram is broadly divided into two regions: in
the first one, at low polydispersity, the phase behaviour mirrors
closely that of a system with a Schulz distribution of rod
lengths~\cite{SpeSol_p2}, exhibiting an I-N coexistence region in the
middle and single-phase I and N regions to the left and right. The
isotropic cloud curve essentially coincides, at these
small values of $\sigma$, with the one found for a Schulz length
distribution~\cite{SpeSol_p2} and indicated by the dashed line in
Fig.~\ref{fig:phased_log_new}. At the other end of the I-N coexistence
region the nematic cloud curve (not shown) is, as expected from the discussion
in Sec.~\ref{sec:log-normal}, essentially unaffected by the presence
of long rods and remains close to its analogue for a Schulz distribution
in the whole range of $\sigma$. 

The second region of the phase diagram, at higher $\sigma$, is more strongly
affected by the relatively large number of long rods in the log-normal
parent distribution, and therefore also by the value of the cutoff
$l_m$.  At the boundary between the two regions, at $\sigma\approx
0.24$, the cloud curve (see Fig.\ref{fig:phased_log_new}) has a
kink. Correspondingly, the properties of the coexisting nematic shadow
phases change discontinuously at this point, from a ``conventional''
nematic phase similar to that found for a Schulz distribution to one
containing more, and more strongly ordered, long rods; see
Figs.~\ref{fig:cloud_rho0} to \ref{fig:shad_length}
below. Intuitively, below the threshold value of $\sigma$ the parent
distribution is narrow enough to keep the number of long rods small,
to the point where phase behaviour is not significantly affected,
while above the threshold the influence of these long rods can become dominant.

Precisely at the kink in the isotropic cloud curve, the isotropic
phase coexists with two different shadow nematic phases. As expected,
we therefore find a three-phase region beginning at this
point. Crossing the isotropic cloud curve from the left, the system
thus first separates into an isotropic phase and an ``unusual''
nematic (N$_1$) with many long rods. Upon increasing density, there is
a narrow interval of three-phase I-N$_1$-N$_2$ coexistence, with N$_2$
being a ``conventional'' nematic phase. Beyond this point, the system
behaves in the usual way, showing an I-N$_2$ coexistence and finally a
single nematic phase. On increasing $\sigma$, however, the three-phase
region eventually closes off and the strict transition between N$_1$
and N$_2$ is replaced by a pronounced crossover in the shape of the
density distribution of the nematic phase; see
Sec.~\ref{sec:the_cutoff_region} below. The crossover line in the
phase diagram essentially continues the three-phase region, and both
are close to the isotropic cloud curve for a Schulz distribution
(dashed line in Fig.~\ref{fig:phased_log_new}), which is
representative also of other parent distributions without fat
tails. We can thus say that it is essentially only to the left of this
line in the phase diagram that the fat tail of the log-normal parent,
and therefore also the value of the cutoff, significantly affect phase
behaviour; one of the resulting effects is of course that, above the
threshold value of $\sigma$, the actual isotropic cloud curve is moved
significantly to the left of its counterpart for a Schulz
distribution. The unaffected region of the phase diagram includes, in
particular, the nematic cloud curve as expected from the discussion
in Sec.~\ref{sec:log-normal}.

The fact that the three-phase region terminates at a finite value of
$\sigma$, where a strict phase coexistence involving two nematic
phases turns into a crossover in the properties of a single nematic phase,
indicates that there must be a nematic-nematic critical point on the
I-N-N phase boundary. However, because the three-phase region itself
is extremely narrow, we have not been able to locate this point
numerically.
 
\subsection{Cloud and shadow curves}
\label{sec:cloud_shadow}

We now discuss in more detail the effect of the long rods on the
isotropic cloud and nematic shadow curves; as shown above, the nematic
cloud and isotropic shadow curves do not need to be considered further
since they are essentially unaffected by the presence of the long
rods.  Fig.~\ref{fig:cloud_rho0} shows the isotropic cloud (main plot)
and nematic shadow (inset) curves, in the $\rho_0$-$\sigma$
representation of plotting polydispersity versus the number densities
$\rho_0$ in the cloud and shadow phases. The kink in the cloud curves
can clearly be seen, as can the associated discontinuity in the shadow
curves.  Upon increasing the cutoff length, the curves also move
towards the predicted limit curve for $l_m\to\infty$. (We defer
a discussion of the numerical results for the actual rate of convergence to
Sec.~\ref{sec:numerical_support}.) The same is true if, as shown in
Fig.~\ref{fig:cloud_shad_rho1}, we use the rescaled volume fractions
$\rho_1$ rather than the number densities to represent the properties
of the cloud and shadow phases. Interestingly, at the discontinuity in
the shadow curves the volume fraction $\rho_1\N $ of the nematic phase
becomes {\em larger} (Fig.~\ref{fig:cloud_shad_rho1}) as $\sigma$ is
increased, while the number density $\rho_0\N$ becomes much smaller
and drops below that of its isotropic counterpart
(Fig.~\ref{fig:cloud_rho0}). This implies an increase in the average
rod length $\langle l \rangle = \rho_1\N/\rho_0\N$ of the nematic
phase, \ie\ a strong fractionation effect which is rather more
pronounced than for a system with a Schulz distribution.
\bfig
\begin{picture}(0,0)%
\includegraphics{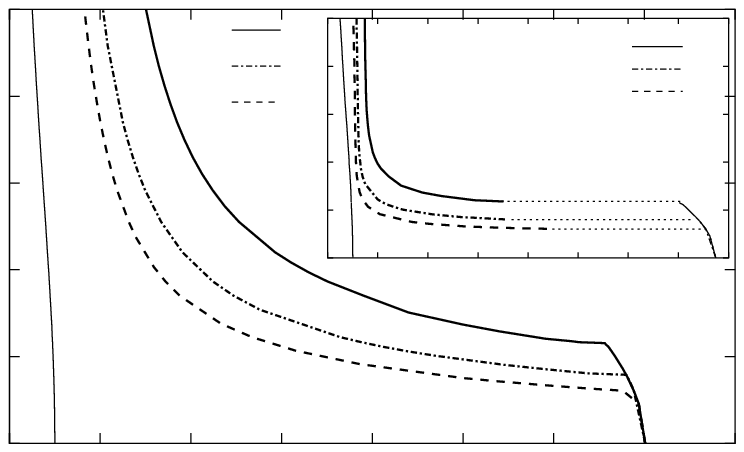}%
\end{picture}%
\setlength{\unitlength}{2565sp}%
\begingroup\makeatletter\ifx\SetFigFont\undefined
% extract first six characters in \fmtname
\def\x#1#2#3#4#5#6#7\relax{\def\x{#1#2#3#4#5#6}}%
\expandafter\x\fmtname xxxxxx\relax \def\y{splain}%
\ifx\x\y   % LaTeX or SliTeX?
\gdef\SetFigFont#1#2#3{%
  \ifnum #1<17\tiny\else \ifnum #1<20\small\else
  \ifnum #1<24\normalsize\else \ifnum #1<29\large\else
  \ifnum #1<34\Large\else \ifnum #1<41\LARGE\else
     \huge\fi\fi\fi\fi\fi\fi
  \csname #3\endcsname}%
\else
\gdef\SetFigFont#1#2#3{\begingroup
  \count@#1\relax \ifnum 25<\count@\count@25\fi
  \def\x{\endgroup\@setsize\SetFigFont{#2pt}}%
  \expandafter\x
    \csname \romannumeral\the\count@ pt\expandafter\endcsname
    \csname @\romannumeral\the\count@ pt\endcsname
  \csname #3\endcsname}%
\fi
\fi\endgroup
\begin{picture}(5802,3727)(301,-3261)
\put(301,-1186){\makebox(0,0)[lb]{\smash{\SetFigFont{8}{9.6}{rm}{\color[rgb]{0,0,0}$\sigma$}%
}}}
\put(5251,-90){\makebox(0,0)[rb]{\smash{\SetFigFont{8}{9.6}{rm}{\color[rgb]{0,0,0}300}%
}}}
\put(5251,-255){\makebox(0,0)[rb]{\smash{\SetFigFont{8}{9.6}{rm}{\color[rgb]{0,0,0}600}%
}}}
\put(2626,-586){\makebox(0,0)[lb]{\smash{\SetFigFont{8}{9.6}{rm}{\color[rgb]{0,0,0}$\sigma$}%
}}}
\put(5251, 75){\makebox(0,0)[rb]{\smash{\SetFigFont{8}{9.6}{rm}{\color[rgb]{0,0,0}100}%
}}}
\put(3526,-3211){\makebox(0,0)[lb]{\smash{\SetFigFont{8}{9.6}{rm}{\color[rgb]{0,0,0}$\rho_0\I$}%
}}}
\put(2309,186){\makebox(0,0)[rb]{\smash{\SetFigFont{8}{9.6}{rm}{\color[rgb]{0,0,0}100}%
}}}
\put(2309,-81){\makebox(0,0)[rb]{\smash{\SetFigFont{8}{9.6}{rm}{\color[rgb]{0,0,0}300}%
}}}
\put(2309,-348){\makebox(0,0)[rb]{\smash{\SetFigFont{8}{9.6}{rm}{\color[rgb]{0,0,0}600}%
}}}
\put(4501,-1751){\makebox(0,0)[lb]{\smash{\SetFigFont{8}{9.6}{rm}{\color[rgb]{0,0,0}$\rho_0\N$}%
}}}
\put(659,-2875){\makebox(0,0)[rb]{\smash{\SetFigFont{8}{9.6}{rm}{\color[rgb]{0,0,0}0}%
}}}
\put(659,-2234){\makebox(0,0)[rb]{\smash{\SetFigFont{8}{9.6}{rm}{\color[rgb]{0,0,0}0.2}%
}}}
\put(659,-1593){\makebox(0,0)[rb]{\smash{\SetFigFont{8}{9.6}{rm}{\color[rgb]{0,0,0}0.4}%
}}}
\put(659,-953){\makebox(0,0)[rb]{\smash{\SetFigFont{8}{9.6}{rm}{\color[rgb]{0,0,0}0.6}%
}}}
\put(659,-312){\makebox(0,0)[rb]{\smash{\SetFigFont{8}{9.6}{rm}{\color[rgb]{0,0,0}0.8}%
}}}
\put(659,329){\makebox(0,0)[rb]{\smash{\SetFigFont{8}{9.6}{rm}{\color[rgb]{0,0,0}1}%
}}}
\put(733,-2999){\makebox(0,0)[b]{\smash{\SetFigFont{8}{9.6}{rm}{\color[rgb]{0,0,0}0}%
}}}
\put(1403,-2999){\makebox(0,0)[b]{\smash{\SetFigFont{8}{9.6}{rm}{\color[rgb]{0,0,0}0.5}%
}}}
\put(2073,-2999){\makebox(0,0)[b]{\smash{\SetFigFont{8}{9.6}{rm}{\color[rgb]{0,0,0}1}%
}}}
\put(2742,-2999){\makebox(0,0)[b]{\smash{\SetFigFont{8}{9.6}{rm}{\color[rgb]{0,0,0}1.5}%
}}}
\put(3412,-2999){\makebox(0,0)[b]{\smash{\SetFigFont{8}{9.6}{rm}{\color[rgb]{0,0,0}2}%
}}}
\put(4082,-2999){\makebox(0,0)[b]{\smash{\SetFigFont{8}{9.6}{rm}{\color[rgb]{0,0,0}2.5}%
}}}
\put(4752,-2999){\makebox(0,0)[b]{\smash{\SetFigFont{8}{9.6}{rm}{\color[rgb]{0,0,0}3}%
}}}
\put(5421,-2999){\makebox(0,0)[b]{\smash{\SetFigFont{8}{9.6}{rm}{\color[rgb]{0,0,0}3.5}%
}}}
\put(6091,-2999){\makebox(0,0)[b]{\smash{\SetFigFont{8}{9.6}{rm}{\color[rgb]{0,0,0}4}%
}}}
\put(3041,-1125){\makebox(0,0)[rb]{\smash{\SetFigFont{8}{9.6}{rm}{\color[rgb]{0,0,0}0.2}%
}}}
\put(3041,-418){\makebox(0,0)[rb]{\smash{\SetFigFont{8}{9.6}{rm}{\color[rgb]{0,0,0}0.6}%
}}}
\put(3041,291){\makebox(0,0)[rb]{\smash{\SetFigFont{8}{9.6}{rm}{\color[rgb]{0,0,0}1}%
}}}
\put(3451,-1591){\makebox(0,0)[b]{\smash{\SetFigFont{8}{9.6}{rm}{\color[rgb]{0,0,0}0.5}%
}}}
\put(4193,-1591){\makebox(0,0)[b]{\smash{\SetFigFont{8}{9.6}{rm}{\color[rgb]{0,0,0}1.5}%
}}}
\put(4932,-1591){\makebox(0,0)[b]{\smash{\SetFigFont{8}{9.6}{rm}{\color[rgb]{0,0,0}2.5}%
}}}
\put(5673,-1591){\makebox(0,0)[b]{\smash{\SetFigFont{8}{9.6}{rm}{\color[rgb]{0,0,0}3.5}%
}}}
\end{picture}
\vspace*{0.2cm}
\caption{Isotropic cloud curve and nematic shadow curve (in the inset)
in the $\rho_0$-$\sigma$ representation, for the different values of
the cutoff $l_m$ shown in the legend. Even for the modest values of
$l_m$ used, the evolution of the curves with increasing $l_m$ is
compatible with the predicted convergence to the same limiting curve
(thin solid line) for both isotropic cloud and nematic shadow.  Note the
discontinuity in each shadow curve, which corresponds to the kink in
the associated cloud curve; here the nematic phase jumps from a
``normal'' phase to one dominated by the long rods. Below the kink or
discontinuity, the cloud and shadow curves are practically
cutoff-independent and essentially identical to those found for a
Schulz distribution.}
\label{fig:cloud_rho0}
\efig
\bfig
\begin{picture}(0,0)%
\includegraphics{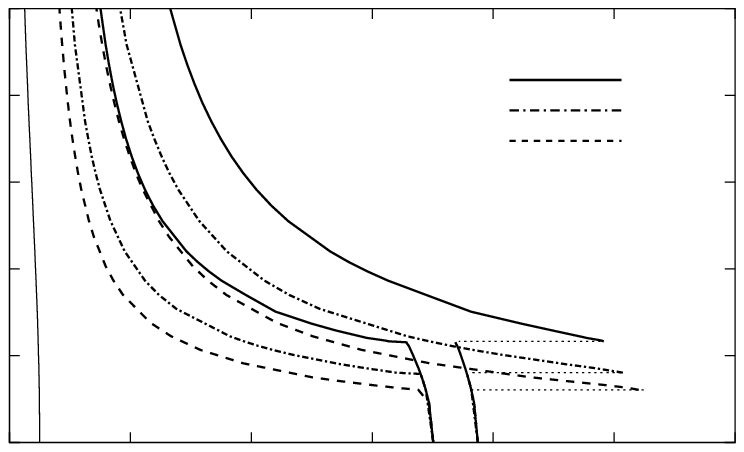}%
\end{picture}%
\setlength{\unitlength}{2565sp}%
\begingroup\makeatletter\ifx\SetFigFont\undefined
% extract first six characters in \fmtname
\def\x#1#2#3#4#5#6#7\relax{\def\x{#1#2#3#4#5#6}}%
\expandafter\x\fmtname xxxxxx\relax \def\y{splain}%
\ifx\x\y   % LaTeX or SliTeX?
\gdef\SetFigFont#1#2#3{%
  \ifnum #1<17\tiny\else \ifnum #1<20\small\else
  \ifnum #1<24\normalsize\else \ifnum #1<29\large\else
  \ifnum #1<34\Large\else \ifnum #1<41\LARGE\else
     \huge\fi\fi\fi\fi\fi\fi
  \csname #3\endcsname}%
\else
\gdef\SetFigFont#1#2#3{\begingroup
  \count@#1\relax \ifnum 25<\count@\count@25\fi
  \def\x{\endgroup\@setsize\SetFigFont{#2pt}}%
  \expandafter\x
    \csname \romannumeral\the\count@ pt\expandafter\endcsname
    \csname @\romannumeral\the\count@ pt\endcsname
  \csname #3\endcsname}%
\fi
\fi\endgroup
\begin{picture}(5802,3881)(301,-3415)
\put(4101,-436){\makebox(0,0)[lb]{\smash{\SetFigFont{8}{9.6}{rm}{\color[rgb]{0,0,0}300}%
}}}
\put(4101,-661){\makebox(0,0)[lb]{\smash{\SetFigFont{8}{9.6}{rm}{\color[rgb]{0,0,0}600}%
}}}
\put(3451,-3361){\makebox(0,0)[lb]{\smash{\SetFigFont{8}{9.6}{rm}{\color[rgb]{0,0,0}$\rho_1=(L_0/D)\ \phi_1$}%
}}}
\put(301,-1261){\makebox(0,0)[lb]{\smash{\SetFigFont{8}{9.6}{rm}{\color[rgb]{0,0,0}$\sigma$}%
}}}
\put(4101,-211){\makebox(0,0)[lb]{\smash{\SetFigFont{8}{9.6}{rm}{\color[rgb]{0,0,0}100}%
}}}
\put(659,-2875){\makebox(0,0)[rb]{\smash{\SetFigFont{8}{9.6}{rm}{\color[rgb]{0,0,0}0}%
}}}
\put(659,-2234){\makebox(0,0)[rb]{\smash{\SetFigFont{8}{9.6}{rm}{\color[rgb]{0,0,0}0.2}%
}}}
\put(659,-1593){\makebox(0,0)[rb]{\smash{\SetFigFont{8}{9.6}{rm}{\color[rgb]{0,0,0}0.4}%
}}}
\put(659,-953){\makebox(0,0)[rb]{\smash{\SetFigFont{8}{9.6}{rm}{\color[rgb]{0,0,0}0.6}%
}}}
\put(659,-312){\makebox(0,0)[rb]{\smash{\SetFigFont{8}{9.6}{rm}{\color[rgb]{0,0,0}0.8}%
}}}
\put(659,329){\makebox(0,0)[rb]{\smash{\SetFigFont{8}{9.6}{rm}{\color[rgb]{0,0,0}1}%
}}}
\put(733,-2999){\makebox(0,0)[b]{\smash{\SetFigFont{8}{9.6}{rm}{\color[rgb]{0,0,0}0}%
}}}
\put(1626,-2999){\makebox(0,0)[b]{\smash{\SetFigFont{8}{9.6}{rm}{\color[rgb]{0,0,0}1}%
}}}
\put(2519,-2999){\makebox(0,0)[b]{\smash{\SetFigFont{8}{9.6}{rm}{\color[rgb]{0,0,0}2}%
}}}
\put(3412,-2999){\makebox(0,0)[b]{\smash{\SetFigFont{8}{9.6}{rm}{\color[rgb]{0,0,0}3}%
}}}
\put(4305,-2999){\makebox(0,0)[b]{\smash{\SetFigFont{8}{9.6}{rm}{\color[rgb]{0,0,0}4}%
}}}
\put(5198,-2999){\makebox(0,0)[b]{\smash{\SetFigFont{8}{9.6}{rm}{\color[rgb]{0,0,0}5}%
}}}
\put(6091,-2999){\makebox(0,0)[b]{\smash{\SetFigFont{8}{9.6}{rm}{\color[rgb]{0,0,0}6}%
}}}
\end{picture}
\caption{Analogue of Fig.~\protect\ref{fig:cloud_rho0}, but in
the $\rho_1$-$\sigma$ representation, \ie\ with (rescaled) volume
fractions $\rho_1$ used instead of number densities $\rho_0$ to
indicate the properties of the phases. Note the large increase in
$\rho_1\N$ at the discontinuity of the nematic shadow curves, which
contrasts with a decrease in $\rho_0\N$ (see
Fig.~\protect\ref{fig:cloud_rho0}). Below
the discontinuity, the results are again very similar to those for a
Schulz distribution.}
\label{fig:cloud_shad_rho1}
\efig
This is shown explicitly in Fig.\ref{fig:shad_length}, where we plot
the average rod length in the nematic shadow phases against the
polydispersity. The discontinuities in the shadow curves for $\rho_1$
and $\rho_0$ give a similar discontinuity here; below it, the results
are very close to those for a Schulz distribution, while for higher
values of $\sigma$ the rod length in the nematic shadow phase is
considerably larger, indicating that the nematic phase has become
enriched in the long rods. (Note that the coexisting isotropic cloud phase
always has average rod length one, since its density distribution is
identical to that of the parent.) As $\sigma$ is increased, the
fractionation first increases, but then reaches a maximum and
decreases again. This can be understood by looking back at
Figs.~\ref{fig:cloud_rho0} and~\ref{fig:cloud_shad_rho1}: the nematic
shadow's number density $\rho_0\N$ initially decreases more quickly
with increasing $\sigma$ than its rod volume fraction $\rho_1\N$,
leading to the observed increase in $\langle
l\rangle=\rho_1\N/\rho_0\N$; but eventually $\rho_0\N$ becomes almost
independent of $\sigma$ and $\langle l \rangle$ then decreases with $\rho_1\N$.
It is intriguing that the maximum fractionation is
reached at roughly the same value of $\sigma$ at which the three-phase
region disappears (compare Fig.~\ref{fig:phased_log_new}), but
we have no explanation for this at present.

\bfig
\begin{picture}(0,0)%
\includegraphics{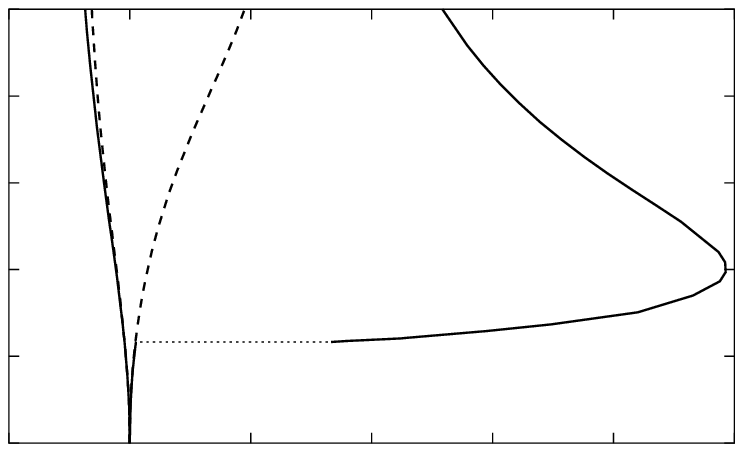}%
\end{picture}%
\setlength{\unitlength}{2693sp}%
\begingroup\makeatletter\ifx\SetFigFont\undefined
% extract first six characters in \fmtname
\def\x#1#2#3#4#5#6#7\relax{\def\x{#1#2#3#4#5#6}}%
\expandafter\x\fmtname xxxxxx\relax \def\y{splain}%
\ifx\x\y   % LaTeX or SliTeX?
\gdef\SetFigFont#1#2#3{%
  \ifnum #1<17\tiny\else \ifnum #1<20\small\else
  \ifnum #1<24\normalsize\else \ifnum #1<29\large\else
  \ifnum #1<34\Large\else \ifnum #1<41\LARGE\else
     \huge\fi\fi\fi\fi\fi\fi
  \csname #3\endcsname}%
\else
\gdef\SetFigFont#1#2#3{\begingroup
  \count@#1\relax \ifnum 25<\count@\count@25\fi
  \def\x{\endgroup\@setsize\SetFigFont{#2pt}}%
  \expandafter\x
    \csname \romannumeral\the\count@ pt\expandafter\endcsname
    \csname @\romannumeral\the\count@ pt\endcsname
  \csname #3\endcsname}%
\fi
\fi\endgroup
\begin{picture}(5670,3635)(143,-3144)
\put(143,-1232){\makebox(0,0)[lb]{\smash{\SetFigFont{9}{10.8}{rm}{\color[rgb]{0,0,0}$\sigma$}%
}}}
\put(2215,-3090){\makebox(0,0)[lb]{\smash{\SetFigFont{9}{10.8}{rm}{\color[rgb]{0,0,0}$\langle l\rangle=\rho_1/\rho_0$}%
}}}
\put(627,-2698){\makebox(0,0)[rb]{\smash{\SetFigFont{9}{10.8}{rm}{\color[rgb]{0,0,0}0}%
}}}
\put(627,-2088){\makebox(0,0)[rb]{\smash{\SetFigFont{9}{10.8}{rm}{\color[rgb]{0,0,0}0.2}%
}}}
\put(627,-1477){\makebox(0,0)[rb]{\smash{\SetFigFont{9}{10.8}{rm}{\color[rgb]{0,0,0}0.4}%
}}}
\put(627,-867){\makebox(0,0)[rb]{\smash{\SetFigFont{9}{10.8}{rm}{\color[rgb]{0,0,0}0.6}%
}}}
\put(627,-257){\makebox(0,0)[rb]{\smash{\SetFigFont{9}{10.8}{rm}{\color[rgb]{0,0,0}0.8}%
}}}
\put(627,354){\makebox(0,0)[rb]{\smash{\SetFigFont{9}{10.8}{rm}{\color[rgb]{0,0,0}1}%
}}}
\put(698,-2816){\makebox(0,0)[b]{\smash{\SetFigFont{9}{10.8}{rm}{\color[rgb]{0,0,0}0}%
}}}
\put(1548,-2816){\makebox(0,0)[b]{\smash{\SetFigFont{9}{10.8}{rm}{\color[rgb]{0,0,0}1}%
}}}
\put(2399,-2816){\makebox(0,0)[b]{\smash{\SetFigFont{9}{10.8}{rm}{\color[rgb]{0,0,0}2}%
}}}
\put(3249,-2816){\makebox(0,0)[b]{\smash{\SetFigFont{9}{10.8}{rm}{\color[rgb]{0,0,0}3}%
}}}
\put(4100,-2816){\makebox(0,0)[b]{\smash{\SetFigFont{9}{10.8}{rm}{\color[rgb]{0,0,0}4}%
}}}
\put(4951,-2816){\makebox(0,0)[b]{\smash{\SetFigFont{9}{10.8}{rm}{\color[rgb]{0,0,0}5}%
}}}
\put(5801,-2816){\makebox(0,0)[b]{\smash{\SetFigFont{9}{10.8}{rm}{\color[rgb]{0,0,0}6}%
}}}
\end{picture}
\caption{The average rod length $\langle l\rangle$ in the nematic and
isotropic shadow phases (solid) at cutoff $l_m=100$, compared with the
same results for a system with a Schulz distribution (dashed). The isotropic
shadow, \ie\ the phase which initially coexists with the nematic when
the density is reduced from large values, is in both cases the one at
lower $\langle l\rangle$; as expected, it is essentially unaffected by
the presence of long rods so that the results for the log-normal and
the Schulz distributions are very similar. The same is true for the
nematic shadow phase below the threshold value of $\sigma$; above,
the average rod length is much larger. The nematic phase here has an
enhanced concentration of the longest rods, leading to pronounced
fractionation.}
\label{fig:shad_length}
\efig
For completeness, Fig.~\ref{fig:shad_length} also shows the average
rod length in the {\em isotropic} shadow; recall that this is
defined as the isotropic
phase which coexists with the nematic cloud phase at the upper limit
(in density) of the phase separation region. As expected, the results
show no pronounced effect from the presence of long rods and are
close to those for a Schulz distribution.

More insight into the properties of the isotropic cloud and nematic
shadow phases can be gained by plotting their density distributions
$\rho(l)$. Fig.~\ref{fig:I_N_distr_both2} shows exemplary results for
$l_m=100$ and a value of $\sigma$ above the discontinuity in the
shadow curve, \ie\ in the region of the phase diagram where the long
rods have a significant effect.
\bfig
\begin{picture}(0,0)%
\includegraphics{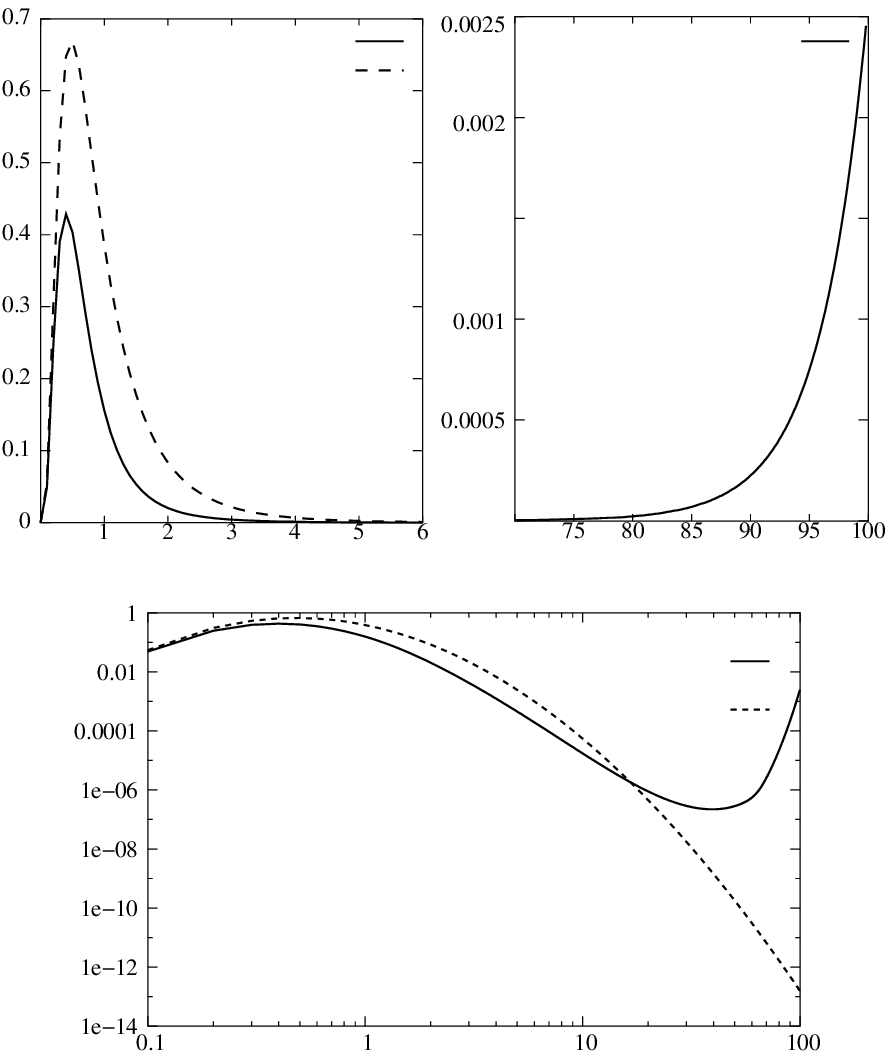}%
\end{picture}%
\setlength{\unitlength}{2447sp}%
\begingroup\makeatletter\ifx\SetFigFont\undefined
% extract first six characters in \fmtname
\def\x#1#2#3#4#5#6#7\relax{\def\x{#1#2#3#4#5#6}}%
\expandafter\x\fmtname xxxxxx\relax \def\y{splain}%
\ifx\x\y   % LaTeX or SliTeX?
\gdef\SetFigFont#1#2#3{%
  \ifnum #1<17\tiny\else \ifnum #1<20\small\else
  \ifnum #1<24\normalsize\else \ifnum #1<29\large\else
  \ifnum #1<34\Large\else \ifnum #1<41\LARGE\else
     \huge\fi\fi\fi\fi\fi\fi
  \csname #3\endcsname}%
\else
\gdef\SetFigFont#1#2#3{\begingroup
  \count@#1\relax \ifnum 25<\count@\count@25\fi
  \def\x{\endgroup\@setsize\SetFigFont{#2pt}}%
  \expandafter\x
    \csname \romannumeral\the\count@ pt\expandafter\endcsname
    \csname @\romannumeral\the\count@ pt\endcsname
  \csname #3\endcsname}%
\fi
\fi\endgroup
\begin{picture}(6860,8260)(719,-9564)
\put(6151,-5611){\makebox(0,0)[lb]{\smash{\SetFigFont{7}{8.4}{rm}{\color[rgb]{0,0,0}$l$}%
}}}
\put(2551,-5686){\makebox(0,0)[lb]{\smash{\SetFigFont{7}{8.4}{rm}{\color[rgb]{0,0,0}$l$}%
}}}
\put(4631,-5315){\makebox(0,0)[b]{\smash{\SetFigFont{7}{8.4}{rm}{\color[rgb]{0,0,0}0}%
}}}
\put(4631,-3019){\makebox(0,0)[rb]{\smash{\SetFigFont{7}{8.4}{rm}{\color[rgb]{0,0,0}0.0015}%
}}}
\put(5811,-6419){\makebox(0,0)[lb]{\smash{\SetFigFont{7}{8.4}{rm}{\color[rgb]{0,0,0}$\rho\N(l)$}%
}}}
\put(5846,-6785){\makebox(0,0)[lb]{\smash{\SetFigFont{7}{8.4}{rm}{\color[rgb]{0,0,0}$\rho\I(l)$}%
}}}
\put(4452,-9515){\makebox(0,0)[lb]{\smash{\SetFigFont{7}{8.4}{rm}{\color[rgb]{0,0,0}$l$}%
}}}
\put(6826,-1636){\makebox(0,0)[rb]{\smash{\SetFigFont{7}{8.4}{rm}{\color[rgb]{0,0,0}$\rho\N(l)$}%
}}}
\put(3376,-1861){\makebox(0,0)[rb]{\smash{\SetFigFont{7}{8.4}{rm}{\color[rgb]{0,0,0}$\rho\I(l)$}%
}}}
\put(3376,-1636){\makebox(0,0)[rb]{\smash{\SetFigFont{7}{8.4}{rm}{\color[rgb]{0,0,0}$\rho\N(l)$}%
}}}
\end{picture}
\caption{Density distribution of the isotropic cloud phase (solid) and the nematic
shadow phase (dashed), at the isotropic cloud point for $l_m=100$ and
$\sigma= 0.63$, in linear (top left and right) and log-log (bottom)
scale. The top right panel shows the region of large $l$,
where the nematic shadow distribution exhibits an exponentially
increasing regime. The bottom panel demonstrates that, given the
very small density of such long rods in the isotropic phase, their enrichment
in the nematic phase is quite dramatic.
}
\label{fig:I_N_distr_both2}
\efig
As anticipated in Sec.~\ref{sec:theory_sum}, the density distribution
of the nematic shadow for rod lengths of order unity has a 
shape similar to that of the isotropic 
cloud phase, while exhibiting an exponential
increase for rod lengths $l$ comparable to the cutoff $l_m$. Note that
the total density in this long rods part of the distribution is much
smaller (by almost two orders of magnitude in the example) than the
overall nematic density; but the fact that the lengths themselves are large
makes this sufficient to give the large average rod lengths
$\langle l \rangle$ that we found above. With increasing cutoff the
peak moves to larger lengths while, as we will see later in more
detail, its weight decreases. This is why, in the limit
$l_m\rightarrow\infty$, the isotropic cloud and nematic shadow phase
become indistinguishable in their number density and rod volume
fraction, \ie\ in their moments $\rho_0$ and $\rho_1$; as a
consequence, the average rod length in the nematic shadow phase also
tends to unity (the value in the parent) in the limit.

Looking ahead to our theoretical treatment below, the fact that the
nematic density distribution has two maxima, one for $l$ of order
unity and a small second peak for $l=l_m$, will allow us to split
integrals over rod lengths into corresponding short and long rod
parts. The log-scale representation of the density distribution in the
nematic cloud phase (Fig.~\ref{fig:I_N_distr_both2}, bottom)
supports the viability of this approach, showing a clear dip of
$\rho\N(l)$ to negligible values between the short and long rod
regimes. The fact that the decay from the second peak in $\rho\N(l)$,
for $l<l_m$, is close to exponential will further simplify matters,
allowing us to replace non-exponential factors in the long rod
integrals by their values $l=l_m$.

Finally, we comment on the strength of orientational order in the
nematic cloud phase. In Sec.~\ref{sec:theory_sum} we anticipated that
in the limit of large cutoff this actually vanishes, as indicated by
the convergence of $\rho_2\N$ to zero. This implies that rods with
lengths $l$ of order unity are, in the limit, orientationally
disordered: Eq.~\eqref{eq:P_l_theory} gives a uniform orientational
distribution when $\rho_2\N l\to 0$. More specifically,
Eq.~\eqref{eq:P_l_theory} implies that only rods with $l$ of order
$1/\rho_2\N$ and greater show significant orientational order, and
$\rho_2\N\to0$ for $l_m\to\infty$ implies that this ``ordering
length'' $1/\rho_2\N$ diverges (while remaining $\ll l_m$; see
Sec.~\ref{sec:scaling}). The convergence of $\rho_2\N$ to zero is
rather slow however, and for the modest values of $l_m$ used so far
$\rho_2\N$ is still of order unity. For example, the nematic shadow
phase at $\sigma=0.5$ and $l_m=100$ has $\rho_2\N\approx 2$ and
therefore noticeable orientational order even for rod lengths $l$ of
order one.

%%%%%%%%%%%%%%%%%%%%%%CUTOFF REGION%%%%%%%%%%%%%%%%%%%%%%%%%%%%%%%%%% 
\subsection{Long rod effects in the coexistence region}
\label{sec:the_cutoff_region}

In Sec.~\ref{sec:phase_diagram}, we reported that the region of the
phase diagram most strongly affected by the long rods is located above
the treshold value of $\sigma$ and between the actual isotropic cloud
curve and that which is obtained for distributions without fat tails,
\eg\ of Schulz type. We now explore this region in more detail,
focussing on the regime of large $\sigma$ where there is no three-phase
I-N-N coexistence but instead a pronounced change in the properties of
the nematic phase as the coexistence region is traversed.

\bfig
\begin{picture}(0,0)%
\includegraphics{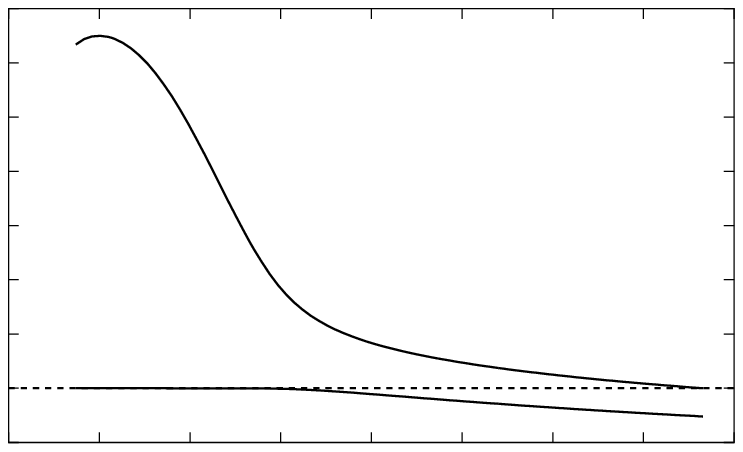}%
\end{picture}%
\setlength{\unitlength}{2565sp}%
\begingroup\makeatletter\ifx\SetFigFont\undefined
% extract first six characters in \fmtname
\def\x#1#2#3#4#5#6#7\relax{\def\x{#1#2#3#4#5#6}}%
\expandafter\x\fmtname xxxxxx\relax \def\y{splain}%
\ifx\x\y   % LaTeX or SliTeX?
\gdef\SetFigFont#1#2#3{%
  \ifnum #1<17\tiny\else \ifnum #1<20\small\else
  \ifnum #1<24\normalsize\else \ifnum #1<29\large\else
  \ifnum #1<34\Large\else \ifnum #1<41\LARGE\else
     \huge\fi\fi\fi\fi\fi\fi
  \csname #3\endcsname}%
\else
\gdef\SetFigFont#1#2#3{\begingroup
  \count@#1\relax \ifnum 25<\count@\count@25\fi
  \def\x{\endgroup\@setsize\SetFigFont{#2pt}}%
  \expandafter\x
    \csname \romannumeral\the\count@ pt\expandafter\endcsname
    \csname @\romannumeral\the\count@ pt\endcsname
  \csname #3\endcsname}%
\fi
\fi\endgroup
\begin{picture}(5887,3881)(226,-3415)
\put(2776,-1486){\makebox(0,0)[lb]{\smash{\SetFigFont{8}{9.6}{rm}{\color[rgb]{0,0,0}N}%
}}}
\put(3076,-2686){\makebox(0,0)[lb]{\smash{\SetFigFont{8}{9.6}{rm}{\color[rgb]{0,0,0}I}%
}}}
\put(3151,-3361){\makebox(0,0)[lb]{\smash{\SetFigFont{8}{9.6}{rm}{\color[rgb]{0,0,0}$\rho_0^{(0)}$}%
}}}
\put(226,-1036){\makebox(0,0)[lb]{\smash{\SetFigFont{8}{9.6}{rm}{\color[rgb]{0,0,0}$\langle l\rangle$}%
}}}
\put(659,-2875){\makebox(0,0)[rb]{\smash{\SetFigFont{8}{9.6}{rm}{\color[rgb]{0,0,0}0.5}%
}}}
\put(659,-2474){\makebox(0,0)[rb]{\smash{\SetFigFont{8}{9.6}{rm}{\color[rgb]{0,0,0}1}%
}}}
\put(659,-2074){\makebox(0,0)[rb]{\smash{\SetFigFont{8}{9.6}{rm}{\color[rgb]{0,0,0}1.5}%
}}}
\put(659,-1673){\makebox(0,0)[rb]{\smash{\SetFigFont{8}{9.6}{rm}{\color[rgb]{0,0,0}2}%
}}}
\put(659,-1273){\makebox(0,0)[rb]{\smash{\SetFigFont{8}{9.6}{rm}{\color[rgb]{0,0,0}2.5}%
}}}
\put(659,-872){\makebox(0,0)[rb]{\smash{\SetFigFont{8}{9.6}{rm}{\color[rgb]{0,0,0}3}%
}}}
\put(659,-472){\makebox(0,0)[rb]{\smash{\SetFigFont{8}{9.6}{rm}{\color[rgb]{0,0,0}3.5}%
}}}
\put(659,-71){\makebox(0,0)[rb]{\smash{\SetFigFont{8}{9.6}{rm}{\color[rgb]{0,0,0}4}%
}}}
\put(659,329){\makebox(0,0)[rb]{\smash{\SetFigFont{8}{9.6}{rm}{\color[rgb]{0,0,0}4.5}%
}}}
\put(733,-2999){\makebox(0,0)[b]{\smash{\SetFigFont{8}{9.6}{rm}{\color[rgb]{0,0,0}0.5}%
}}}
\put(1403,-2999){\makebox(0,0)[b]{\smash{\SetFigFont{8}{9.6}{rm}{\color[rgb]{0,0,0}1}%
}}}
\put(2073,-2999){\makebox(0,0)[b]{\smash{\SetFigFont{8}{9.6}{rm}{\color[rgb]{0,0,0}1.5}%
}}}
\put(2742,-2999){\makebox(0,0)[b]{\smash{\SetFigFont{8}{9.6}{rm}{\color[rgb]{0,0,0}2}%
}}}
\put(3412,-2999){\makebox(0,0)[b]{\smash{\SetFigFont{8}{9.6}{rm}{\color[rgb]{0,0,0}2.5}%
}}}
\put(4082,-2999){\makebox(0,0)[b]{\smash{\SetFigFont{8}{9.6}{rm}{\color[rgb]{0,0,0}3}%
}}}
\put(4752,-2999){\makebox(0,0)[b]{\smash{\SetFigFont{8}{9.6}{rm}{\color[rgb]{0,0,0}3.5}%
}}}
\put(5421,-2999){\makebox(0,0)[b]{\smash{\SetFigFont{8}{9.6}{rm}{\color[rgb]{0,0,0}4}%
}}}
\put(6091,-2999){\makebox(0,0)[b]{\smash{\SetFigFont{8}{9.6}{rm}{\color[rgb]{0,0,0}4.5}%
}}}
\end{picture}
\vspace*{0.5cm}
\caption{Average rod length in the isotropic and nematic phase across
the coexistence region at $\sigma=0.7$ and $l_m=100$. Notice the
strong fractionation in the first, ``long rod-dominated'', part of the
coexistence region up to parent densities around $\parent_0=2$, while
above the behaviour resembles that for a parent with \eg\ a Schulz
distribution.  Notice also the non-monotonicity of the average rod
length in the nematic phase, which is consistent with our theory for
the large-cutoff limit (see text).}
\label{fig:length_07_100}
\efig
While at the isotropic cloud point it is only the longer rods in
the nematic phase that exhibit noticeable orientational order (at
least for large cutoff $l_m$; see end of previous section), we
expect also the shorter rods to become ordered eventually as density
is increased. As the effect of the longest rods on the nematic phase
diminishes, fractionation effects are then expected to become less
pronounced, leading to a reduced average rod length in the nematic.  A
plot of the average rod lengths in the isotropic and nematic phases
against the parent density across the coexistence region
(Fig.~\ref{fig:length_07_100}) shows this effect clearly. At parent
densities $\parent_0\approx 2$ we see a crossover, where the average
rod length in the nematic drops to the values typical for phase
separation from a parent distribution without a fat tail; as expected,
the crossover density is close to the isotropic cloud point
($\parent_0\simeq 2.1 $) for a Schulz distribution with the
same polydispersity. Notice that the average rod length in the nematic
phase, $\langle l\rangle\N$, is non-monotonic in parent density,
initially showing an increase when moving away from the isotropic
cloud point but then decreasing again. This is consistent with our
theory for the large cutoff-limit, which predicts that for
$l_m\to\infty$, $\langle l \rangle\N=\langle l \rangle\I$ ($=1$ with
our convention for the parent's average rod length) at the isotropic
cloud point. On increasing the density, $\langle l\rangle\N$ must then
first increase as the system begins to fractionate, before reducing
again to unity as the nematic cloud point is approached.

\bfig
\begin{picture}(0,0)%
\includegraphics{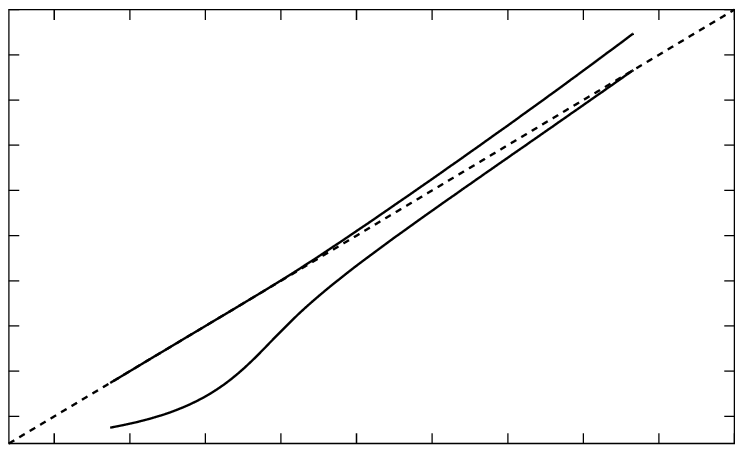}%
\end{picture}%
\setlength{\unitlength}{2565sp}%
\begingroup\makeatletter\ifx\SetFigFont\undefined
% extract first six characters in \fmtname
\def\x#1#2#3#4#5#6#7\relax{\def\x{#1#2#3#4#5#6}}%
\expandafter\x\fmtname xxxxxx\relax \def\y{splain}%
\ifx\x\y   % LaTeX or SliTeX?
\gdef\SetFigFont#1#2#3{%
  \ifnum #1<17\tiny\else \ifnum #1<20\small\else
  \ifnum #1<24\normalsize\else \ifnum #1<29\large\else
  \ifnum #1<34\Large\else \ifnum #1<41\LARGE\else
     \huge\fi\fi\fi\fi\fi\fi
  \csname #3\endcsname}%
\else
\gdef\SetFigFont#1#2#3{\begingroup
  \count@#1\relax \ifnum 25<\count@\count@25\fi
  \def\x{\endgroup\@setsize\SetFigFont{#2pt}}%
  \expandafter\x
    \csname \romannumeral\the\count@ pt\expandafter\endcsname
    \csname @\romannumeral\the\count@ pt\endcsname
  \csname #3\endcsname}%
\fi
\fi\endgroup
\begin{picture}(5454,3881)(659,-3415)
\put(3451,-886){\makebox(0,0)[lb]{\smash{\SetFigFont{8}{9.6}{rm}{\color[rgb]{0,0,0}$\rho_0\I$}%
}}}
\put(3001,-2086){\makebox(0,0)[lb]{\smash{\SetFigFont{8}{9.6}{rm}{\color[rgb]{0,0,0}$\rho_0\N$}%
}}}
\put(3001,-3361){\makebox(0,0)[lb]{\smash{\SetFigFont{8}{9.6}{rm}{\color[rgb]{0,0,0}$\rho_0^{(0)}$}%
}}}
\put(659,-2675){\makebox(0,0)[rb]{\smash{\SetFigFont{8}{9.6}{rm}{\color[rgb]{0,0,0}0.5}%
}}}
\put(659,-2341){\makebox(0,0)[rb]{\smash{\SetFigFont{8}{9.6}{rm}{\color[rgb]{0,0,0}1}%
}}}
\put(659,-2007){\makebox(0,0)[rb]{\smash{\SetFigFont{8}{9.6}{rm}{\color[rgb]{0,0,0}1.5}%
}}}
\put(659,-1674){\makebox(0,0)[rb]{\smash{\SetFigFont{8}{9.6}{rm}{\color[rgb]{0,0,0}2}%
}}}
\put(659,-1340){\makebox(0,0)[rb]{\smash{\SetFigFont{8}{9.6}{rm}{\color[rgb]{0,0,0}2.5}%
}}}
\put(659,-1006){\makebox(0,0)[rb]{\smash{\SetFigFont{8}{9.6}{rm}{\color[rgb]{0,0,0}3}%
}}}
\put(659,-672){\makebox(0,0)[rb]{\smash{\SetFigFont{8}{9.6}{rm}{\color[rgb]{0,0,0}3.5}%
}}}
\put(659,-339){\makebox(0,0)[rb]{\smash{\SetFigFont{8}{9.6}{rm}{\color[rgb]{0,0,0}4}%
}}}
\put(659, -5){\makebox(0,0)[rb]{\smash{\SetFigFont{8}{9.6}{rm}{\color[rgb]{0,0,0}4.5}%
}}}
\put(659,329){\makebox(0,0)[rb]{\smash{\SetFigFont{8}{9.6}{rm}{\color[rgb]{0,0,0}5}%
}}}
\put(1068,-2999){\makebox(0,0)[b]{\smash{\SetFigFont{8}{9.6}{rm}{\color[rgb]{0,0,0}0.5}%
}}}
\put(1626,-2999){\makebox(0,0)[b]{\smash{\SetFigFont{8}{9.6}{rm}{\color[rgb]{0,0,0}1}%
}}}
\put(2184,-2999){\makebox(0,0)[b]{\smash{\SetFigFont{8}{9.6}{rm}{\color[rgb]{0,0,0}1.5}%
}}}
\put(2742,-2999){\makebox(0,0)[b]{\smash{\SetFigFont{8}{9.6}{rm}{\color[rgb]{0,0,0}2}%
}}}
\put(3300,-2999){\makebox(0,0)[b]{\smash{\SetFigFont{8}{9.6}{rm}{\color[rgb]{0,0,0}2.5}%
}}}
\put(3858,-2999){\makebox(0,0)[b]{\smash{\SetFigFont{8}{9.6}{rm}{\color[rgb]{0,0,0}3}%
}}}
\put(4417,-2999){\makebox(0,0)[b]{\smash{\SetFigFont{8}{9.6}{rm}{\color[rgb]{0,0,0}3.5}%
}}}
\put(4975,-2999){\makebox(0,0)[b]{\smash{\SetFigFont{8}{9.6}{rm}{\color[rgb]{0,0,0}4}%
}}}
\put(5533,-2999){\makebox(0,0)[b]{\smash{\SetFigFont{8}{9.6}{rm}{\color[rgb]{0,0,0}4.5}%
}}}
\put(6091,-2999){\makebox(0,0)[b]{\smash{\SetFigFont{8}{9.6}{rm}{\color[rgb]{0,0,0}5}%
}}}
\end{picture}
\vspace*{0.5cm}
\caption{Number density of the isotropic and nematic phases across the
coexistence region for $\sigma=0.7$ and $l_m=100$, plotted against the
parent density. Notice that in the long rod-dominated region below
$\parent_0\approx 2$, the density $\rho_0\I$ of the isotropic phase is
almost identical to that of the parent (dashed line). In the same
regime, the nematic density $\rho_0\N$ shows a pronounced increase,
before crossing over to the more gradual variation typical of systems
without fat-tailed length distributions.}
\label{fig:rho_07_100}
\efig  
The variation of the density of the two phases across the coexistence
region (Fig.~\ref{fig:rho_07_100}) similarly shows the crossover
between two different regimes for the nematic phase: in the
low-density regime, the density of the nematic increases rather
quickly with the parent density, before crossing over to the more
gradual variation that is familiar from systems with length
distributions without fat tails.  Interestingly, the density of the
isotropic phase is essentially identical to that of the parent
throughout the low-density region where the phase separation behaviour
is dominated by the long rods.  Since the overall density of the
system must equal that of the parent, the fractional system volume
occupied by the nematic phase must then be very small. This is indeed
what we find, and the representation in Fig.~\ref{fig:vol_07_100} of
the phase behaviour provides probably the clearest visual
demonstration of the crossover between the low-density regime
dominated by long rods and the region of more conventional phase
behaviour at higher densities.

\bfig
\begin{picture}(0,0)%
\includegraphics{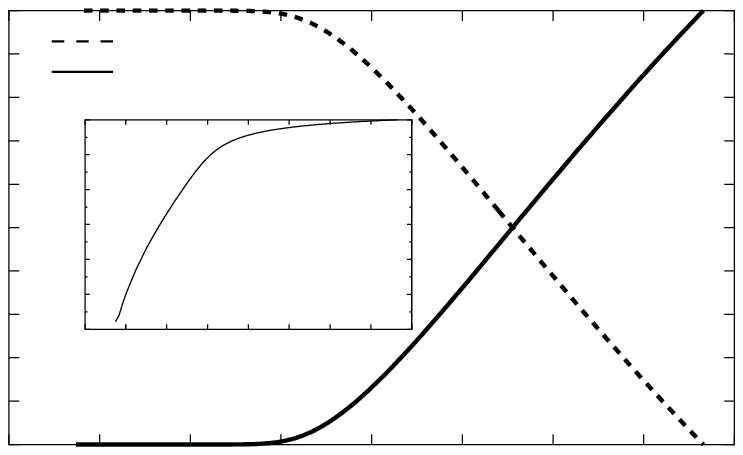}%
\end{picture}%
\setlength{\unitlength}{2565sp}%
\begingroup\makeatletter\ifx\SetFigFont\undefined
% extract first six characters in \fmtname
\def\x#1#2#3#4#5#6#7\relax{\def\x{#1#2#3#4#5#6}}%
\expandafter\x\fmtname xxxxxx\relax \def\y{splain}%
\ifx\x\y   % LaTeX or SliTeX?
\gdef\SetFigFont#1#2#3{%
  \ifnum #1<17\tiny\else \ifnum #1<20\small\else
  \ifnum #1<24\normalsize\else \ifnum #1<29\large\else
  \ifnum #1<34\Large\else \ifnum #1<41\LARGE\else
     \huge\fi\fi\fi\fi\fi\fi
  \csname #3\endcsname}%
\else
\gdef\SetFigFont#1#2#3{\begingroup
  \count@#1\relax \ifnum 25<\count@\count@25\fi
  \def\x{\endgroup\@setsize\SetFigFont{#2pt}}%
  \expandafter\x
    \csname \romannumeral\the\count@ pt\expandafter\endcsname
    \csname @\romannumeral\the\count@ pt\endcsname
  \csname #3\endcsname}%
\fi
\fi\endgroup
\begin{picture}(5444,3806)(659,-3340)
\put(3151,-3286){\makebox(0,0)[lb]{\smash{\SetFigFont{8}{9.6}{rm}{\color[rgb]{0,0,0}$\rho_0^{(0)}$}%
}}}
\put(1651, 89){\makebox(0,0)[lb]{\smash{\SetFigFont{8}{9.6}{rm}{\color[rgb]{0,0,0}$v\I$}%
}}}
\put(1651,-136){\makebox(0,0)[lb]{\smash{\SetFigFont{8}{9.6}{rm}{\color[rgb]{0,0,0}$v\N$}%
}}}
\put(1260,-703){\makebox(0,0)[rb]{\smash{\SetFigFont{8}{9.6}{rm}{\color[rgb]{0,0,0}1e-02}%
}}}
\put(1260,-1739){\makebox(0,0)[rb]{\smash{\SetFigFont{8}{9.6}{rm}{\color[rgb]{0,0,0}1e-10}%
}}}
\put(1260,-1218){\makebox(0,0)[rb]{\smash{\SetFigFont{8}{9.6}{rm}{\color[rgb]{0,0,0}1e-06}%
}}}
\put(659,-2875){\makebox(0,0)[rb]{\smash{\SetFigFont{8}{9.6}{rm}{\color[rgb]{0,0,0}0}%
}}}
\put(659,-2555){\makebox(0,0)[rb]{\smash{\SetFigFont{8}{9.6}{rm}{\color[rgb]{0,0,0}0.1}%
}}}
\put(659,-2234){\makebox(0,0)[rb]{\smash{\SetFigFont{8}{9.6}{rm}{\color[rgb]{0,0,0}0.2}%
}}}
\put(659,-1914){\makebox(0,0)[rb]{\smash{\SetFigFont{8}{9.6}{rm}{\color[rgb]{0,0,0}0.3}%
}}}
\put(659,-1593){\makebox(0,0)[rb]{\smash{\SetFigFont{8}{9.6}{rm}{\color[rgb]{0,0,0}0.4}%
}}}
\put(659,-953){\makebox(0,0)[rb]{\smash{\SetFigFont{8}{9.6}{rm}{\color[rgb]{0,0,0}0.6}%
}}}
\put(659,-632){\makebox(0,0)[rb]{\smash{\SetFigFont{8}{9.6}{rm}{\color[rgb]{0,0,0}0.7}%
}}}
\put(659,-312){\makebox(0,0)[rb]{\smash{\SetFigFont{8}{9.6}{rm}{\color[rgb]{0,0,0}0.8}%
}}}
\put(659,  9){\makebox(0,0)[rb]{\smash{\SetFigFont{8}{9.6}{rm}{\color[rgb]{0,0,0}0.9}%
}}}
\put(659,329){\makebox(0,0)[rb]{\smash{\SetFigFont{8}{9.6}{rm}{\color[rgb]{0,0,0}1}%
}}}
\put(733,-2999){\makebox(0,0)[b]{\smash{\SetFigFont{8}{9.6}{rm}{\color[rgb]{0,0,0}0.5}%
}}}
\put(1403,-2999){\makebox(0,0)[b]{\smash{\SetFigFont{8}{9.6}{rm}{\color[rgb]{0,0,0}1}%
}}}
\put(2073,-2999){\makebox(0,0)[b]{\smash{\SetFigFont{8}{9.6}{rm}{\color[rgb]{0,0,0}1.5}%
}}}
\put(2742,-2999){\makebox(0,0)[b]{\smash{\SetFigFont{8}{9.6}{rm}{\color[rgb]{0,0,0}2}%
}}}
\put(3412,-2999){\makebox(0,0)[b]{\smash{\SetFigFont{8}{9.6}{rm}{\color[rgb]{0,0,0}2.5}%
}}}
\put(4082,-2999){\makebox(0,0)[b]{\smash{\SetFigFont{8}{9.6}{rm}{\color[rgb]{0,0,0}3}%
}}}
\put(4752,-2999){\makebox(0,0)[b]{\smash{\SetFigFont{8}{9.6}{rm}{\color[rgb]{0,0,0}3.5}%
}}}
\put(5421,-2999){\makebox(0,0)[b]{\smash{\SetFigFont{8}{9.6}{rm}{\color[rgb]{0,0,0}4}%
}}}
\put(6091,-2999){\makebox(0,0)[b]{\smash{\SetFigFont{8}{9.6}{rm}{\color[rgb]{0,0,0}4.5}%
}}}
\put(659,-1273){\makebox(0,0)[rb]{\smash{\SetFigFont{8}{9.6}{rm}{\color[rgb]{0,0,0}0.5}%
}}}
\put(2200,-2071){\makebox(0,0)[b]{\smash{\SetFigFont{8}{9.6}{rm}{\color[rgb]{0,0,0}2}%
}}}
\put(2803,-2071){\makebox(0,0)[b]{\smash{\SetFigFont{8}{9.6}{rm}{\color[rgb]{0,0,0}3}%
}}}
\put(3407,-2071){\makebox(0,0)[b]{\smash{\SetFigFont{8}{9.6}{rm}{\color[rgb]{0,0,0}4}%
}}}
\put(1596,-2071){\makebox(0,0)[b]{\smash{\SetFigFont{8}{9.6}{rm}{\color[rgb]{0,0,0}1}%
}}}
\end{picture}
\vspace*{0.5cm}
\caption{Fractions of system volume occupied by the isotropic (dashed)
and nematic (solid) phases, plotted against parent density across the
coexistence region for $\sigma=0.7$ and $l_m=100$.  On a logarithmic
scale (inset) it is clear the in the regime where the phase separation
is dominated by the long rods, the fractional volume of the nematic is
extremely small.}
\label{fig:vol_07_100}
\efig

The most detailed information about the phase separation is obtained
by analysing the density distributions in the coexisting
phases. Focussing on the low-density part of the coexistence region
that is dominated by the effect of the long rods, we find as expected
from Figs.~\ref{fig:rho_07_100} and~\ref{fig:vol_07_100}
that the density distribution of the
isotropic phase (not shown) is essentially identical to that of the
parent, except for a reduction in the already small
density of long rods.  For the
nematic phase, we saw in Fig.~\ref{fig:I_N_distr_both2} that
$\rho\N(l)$ has a peak at $l=l_m$ at the isotropic cloud point.  When
we move into the coexistence region by increasing the parent density
(Fig.~\ref{fig:N_distr_dens_both}), the density distribution acquires
a bimodal shape as the second peak gradually moves to smaller values
of $l$. Eventually, at the point where the system crosses over to
conventional phase behaviour, this second peak merges with the main
peak at $l$ of order unity. Before this happens, the nematic phase
remains significantly enriched, compared to the parent in some of the longer rods. 
Since the overall density distribution must be preserved
in phase separation (see Eq.~\eqref{eq:lever_rule_theory}), this again
makes plausible that the fractional system volume occupied by nematic
phase has to be very small in this regime: there simply are not enough
long rods to allow formation of a significant amount of a nematic
phase that contains a much higher proportion of such long rods.

\bfig 
\begin{picture}(0,0)%
\includegraphics{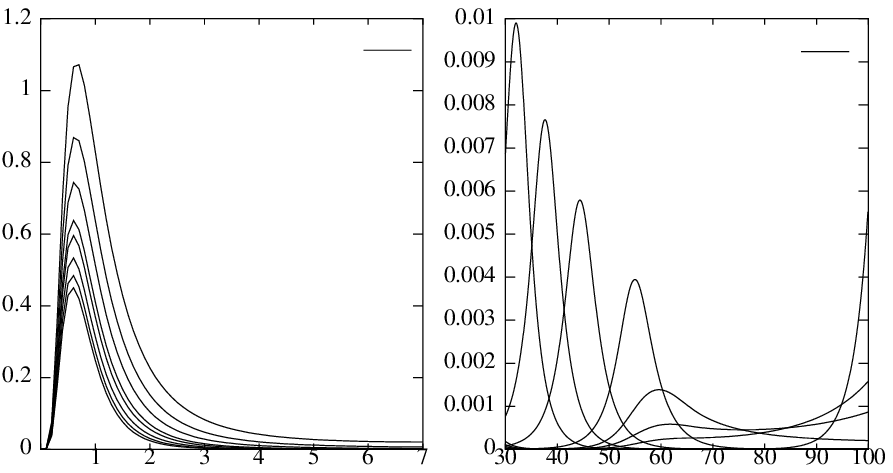}%
\end{picture}%
\setlength{\unitlength}{2447sp}%
\begingroup\makeatletter\ifx\SetFigFont\undefined
% extract first six characters in \fmtname
\def\x#1#2#3#4#5#6#7\relax{\def\x{#1#2#3#4#5#6}}%
\expandafter\x\fmtname xxxxxx\relax \def\y{splain}%
\ifx\x\y   % LaTeX or SliTeX?
\gdef\SetFigFont#1#2#3{%
  \ifnum #1<17\tiny\else \ifnum #1<20\small\else
  \ifnum #1<24\normalsize\else \ifnum #1<29\large\else
  \ifnum #1<34\Large\else \ifnum #1<41\LARGE\else
     \huge\fi\fi\fi\fi\fi\fi
  \csname #3\endcsname}%
\else
\gdef\SetFigFont#1#2#3{\begingroup
  \count@#1\relax \ifnum 25<\count@\count@25\fi
  \def\x{\endgroup\@setsize\SetFigFont{#2pt}}%
  \expandafter\x
    \csname \romannumeral\the\count@ pt\expandafter\endcsname
    \csname @\romannumeral\the\count@ pt\endcsname
  \csname #3\endcsname}%
\fi
\fi\endgroup
\begin{picture}(6860,3839)(419,-5729)
\put(2101,-5686){\makebox(0,0)[lb]{\smash{\SetFigFont{7}{8.4}{rm}{\color[rgb]{0,0,0}$l$}%
}}}
\put(5551,-5686){\makebox(0,0)[lb]{\smash{\SetFigFont{7}{8.4}{rm}{\color[rgb]{0,0,0}$l$}%
}}}
\put(3158,-2273){\makebox(0,0)[rb]{\smash{\SetFigFont{7}{8.4}{rm}{\color[rgb]{0,0,0}$\rho\N(l)$}%
}}}
\put(6546,-2283){\makebox(0,0)[rb]{\smash{\SetFigFont{7}{8.4}{rm}{\color[rgb]{0,0,0}$\rho\N(l)$}%
}}}
\end{picture}
\caption{Plot of the density distribution of the nematic phase for
short (left) and long (right) rods, for a range of parent densities
across the coexistence region at $\sigma=0.28$ and $l_m=100$.  With
increasing parent density, \ie\ when moving away from the isotropic
cloud point, the first (left) peak of the distribution grows and
approaches the shape of the parent.
The second peak at $l_m$, on the other hand, first decreases in
height and then gives way to a maximum at $l<l_m$ which moves towards
smaller $l$.  The crossover to the region in the phase diagram where
long rods no longer significantly affect the phase behaviour takes
place when this maximum merges with the main maximum on the left.}
\label{fig:N_distr_dens_both}
\efig    

Finally, we may ask how the properties of the coexisting phases depend
on the cutoff $l_m$, at some fixed point in the region of the
phase diagram where
the long rods significantly affect the behaviour (\ie\ to the left of
the isotropic cloud curve for rod length distributions without fat
tails). Naively, one might expect that the cutoff would play an
important role not just at the isotropic cloud point, but throughout
this long rod-dominated part of the phase diagram. However, this is
not so. It turns out, for example, that the peak in $\rho\N(l)$ at
large $l$ (but below the cutoff) does not change significantly
as $l_m$ increases and eventually stabilizes. The same applies to
all other properties of the coexisting phases, including the
fractional volume $v\N$ of the nematic phase; the latter remains
nonzero (though very small) as $l_m$ becomes large. Even while the
isotropic cloud curve moves towards the limiting curve for infinite
cutoff (Eq.~\eqref{eq:CPC_rho_0}), widening the phase coexistence
region, the phase separation behaviour in
the region beyond the cloud point therefore eventually becomes
independent of $l_m$. 

%%%%%%%%%%%%%%%%%%%%THEORY%%%%%%%%%%%%%%%%%%%%%%%%%%%%%%%%%
\section{Theory for isotropic cloud point}
\label{sec:limit}

Having discussed the phase behaviour of systems with a log-normal
parent distribution with a finite length cutoff $l_m$, as obtained
from our numerical calculations, we now turn to the limit of large
$l_m$ where the onset of nematic ordering, \ie\ the properties of the
isotropic cloud and nematic shadow phases, are amenable to theoretical
treatment. Even though any physical system will necessarily have a
finite largest rod length and therefore a finite $l_m$, the limiting
case $l_m\to\infty$ is useful in understanding some of the physics
that we have described in the previous sections. In particular, we can
extract a large amount of information about the characteristics of the
coexisting phases, such as the scaling of all the densities of the
coexisting phases with $l_m$, which can be compared directly with our
numerical results. The cloud and shadow curves can be deduced
explicitly in the limit $l_m\to\infty$ and actually coincide. The
moment density $\rho_2\N$ of the nematic shadow phase vanishes in the
limit. Since $\rho_2\N$ specifies the degree of orientational
ordering, one may doubt whether this solution can justifiably be
referred to as a nematic phase. However, recall from
Eq.~\eqref{eq:rho2_S(l)} that $\rho_2\N$ is just $\rho_0\N\langle
S(l)l\rangle$, where the average is taken over the normalized length
distribution in the nematic, $P\N(l)=\rho\N(l)/\rho_0\N$. A small
value of $\rho_2\N$ then implies that $S(l)$, the orientational order
parameter for rods of length $l$ defined in Eq.~\eqref{eq:S(l)}, must
be close to zero wherever $P\N(l)$ is non-vanishing. 
In regions where $P\N(l)$ has
little weight, on the other hand, the value of $S(l)$ can
be close to one. This is exactly what happens in the nematic shadow
phase in the limit $l_m\to\infty$. For $l$ of order unity
$P\N(l)$ has nonzero weight as can be seen from the plot of
$\rho\N(l)=\rho_0\N P\N(l)$ in
Fig.~\ref{fig:I_N_distr_both2}. Correspondingly, 
$S(l)\to 0$ in this regime of rod lengths, implying that the short rods are
orientationally disordered. For the longest
rods with $l$ of order $l_m$, on the other hand,
$P\N(l)$ has negligible weight for $l_m\to\infty$, and we will find
that these rods are strongly ordered, with $S(l)\to1$.
The contributions to $\rho_2\N$ from both
regimes vanish in the limit, although nontrivially they are of the
same order; the fact that the length distribution  has much more
weight on short rods is compensated 
by the smallness of $S(l)$ there.

In the next sections we will establish, by a process of elimination, the
overall limiting behaviour of the relevant quantities as $l_m$ becomes
large. Once we have derived the system of equations to be solved for
finite $l_m$, we will decompose all integrals over $l$ into two parts,
one for lengths of order unity, $l=0\ldots\tilde l$ and one for the
large lengths $l=\tilde l\ldots l_m$. Here $\tilde l$ is chosen such
that $l_m\gg\tilde l\gg 1$; formally, we first take $l_m\to\infty$ and
then $\tilde l\to\infty$. The decomposition into the two parts allows
us to simplify all integrals, making use of the fact that the short
rods are orientationally disordered while the long ones are strongly
ordered.  From this we eventually derive the solution for the
isotropic cloud point and the scaling of all quantities with $l_m$.

%%%%%%%%%%%%%%%%%%%%BETA LIMIT%%%%%%%%%%%%%%%%%%%%%%%%%%%%%%
\subsection{Asymptotic behaviour of $\beta$ and $\rho_2\N$}

Let us start our analysis with a discussion of the behaviour, in the
limit $l_m\rightarrow\infty$, of the key quantity $\beta$; from
Eq.~\eqref{eq:nematic_distr} it follows that its value has a
strong effect on the properties of the nematic shadow
phase. Explicitly, we have from Eq.~\eqref{eq:nematic_distr}
that the moments of the nematic shadow phase are
\begin{eqnarray}
%\label{eq:moments_nematic}
\rho_0\N  &=&\rho_0\totint\normparent(l)\ e^{\beta
l}e^{lf(\theta)}\label{eq:rho0N}\\
%\totint\parent(l)\ e^{\beta l}e^{lf(\theta)}=
\rho_1\N  &=&\rho_0\totint\normparent(l)\ l\ e^{\beta l}\
e^{lf(\theta)}\label{eq:rho1N}\\
%\totint\parent(l)\ l\ e^{\beta l}\ e^{lf(\theta)}=
\rho_2\N  &=&\rho_0\totint\normparent(l)\ l\ e^{\beta l}\
\Ptwo(\cos\theta)\ e^{lf(\theta)}\label{eq:rho2N}
%\totint\parent(l)\ l\ e^{\beta l}\ \Ptwo(\cos\theta)\e^{lf(\theta)}
\end{eqnarray}
where we have defined $f(\theta)$ as
\beq\label{eq:f} 
f(\theta)=c_2\rho_2\N (\Ptwo-1)
\eeq
and have used the fact that the isotropic phase coincides with the
parent distribution $\parent(l)$ at the isotropic cloud point;
$\rho_0=\lint\parent(l)$ is the parent density (we drop the
superscript on $\parent_0$ for brevity). An important constraint on
the dependence of $\beta$ on $l_m$ is that all the above moment
densities must remain finite as $l_m\to\infty$. This follows from the
equality of the osmotic pressure~\eqref{eq:Pi_theory}, which gives
\beq\label{eq:osmotic_pressure}
\rho_0+\frac{1}{2}c_1\rho_0^2=\rho_0\N +\frac{1}{2}c_1\left(\rho_1\N
\right)^2-\frac{1}{2}c_2\left(\rho_2\N \right)^2
\eeq
On the l.h.s.\ we have used again that the isotropic cloud phase
coincides with the parent, giving $\rho_0\I=\rho_0$ and also
$\rho_1\I=\rho_0\I=\rho_0$ since the average rod length in the parent
is unity. All terms on the l.h.s.\ of Eq.~\eqref{eq:Pi_theory} are
finite. The r.h.s.\ is $\geq \rho_0\N + (c_1-c_2)(\rho_1\N)^2/2$ since
$\rho_2\N\leq \rho_1\N$ from Eqs.\eqq{eq:rho1N}{eq:rho2N}. Using
$c_1>c_2$ it follows that both terms are positive, so that $\rho_0\N$
and $\rho_1\N$, and therefore also $\rho_2\N$, must remain finite as
claimed. As in Sec.~\ref{sec:log-normal}, we can now deduce that a
length cutoff $l_m$ must be imposed on all integrals over $l$ whenever
the parent distribution is less than exponentially decaying at large
$l$. This follows because $\beta$ should be positive in order to have
a greater rod volume fraction in the nematic phase than in the
isotropic, \ie\ $\rho_1\N\geq\rho_1\I$. Without a cutoff on $l$, such
a positive value of $\beta$ in Eqs.\eqq{eq:rho0N}{eq:rho1N} would
lead to divergent integrals for $\rho_0\N$ and $\rho_1\N$, violating
the constraint that these moment densities must be finite. (The
angular integrals cannot counteract the effect of the exponential
factor $\exp(\beta l)$, since they vary at most as power laws with
$l$.) The cutoff $l_m$ will be assumed large and later taken to
infinity to study the limiting behaviour for large cutoffs. Initially,
we develop the theory for general parent distributions with fat tails,
but then eventually specialize to the log-normal case in order to
compare with our numerical results for finite
cutoffs~\cite{norm_parent}. 

We are now in a position to narrow down the limiting behaviour of
$\beta$ as $l_m\to\infty$.  With $\beta$ being positive, there are
three possibilities: $\beta\to\infty$, $\beta\to$ const.\ and
$\beta\to 0$.  The first case can be excluded from
Eq.~\eqref{eq:beta_def}, which using $\rho_1\I=\rho_0$
becomes
\beq
\label{eq:beta}
\beta=-c_1\left(\rho_1\N -\rho_0\right)+c_2\rho_2\N \eeq We have
already established that all terms on the r.h.s.\ of this equation
remain finite, so that $\beta$ cannot diverge. In the second case,
$\beta\rightarrow$ const.$>0$, the integral in Eq.~\eqref{eq:rho0N}
for $\rho_0\N$ will diverge as $l_m\to\infty$, giving
$\rho_0\N/\rho_0\to\infty$ and therefore $\rho_0\to 0$ to keep
$\rho_0\N$ finite. But then the l.h.s.\ of the osmotic pressure
equality~\eqref{eq:osmotic_pressure} becomes $\simeq \rho_0$, while
the r.h.s.\ is $\geq \rho_0\N + (c_1-c_2)(\rho_1\N)^2/2\geq \rho_0\N$
as explained above; the pressure equality can therefore not be
satisfied since $\rho_0\N\gg \rho_0$. We therefore conclude that the
only possible limit behaviour is $\beta\to 0$ for $l_m\to\infty$.
This means that in the ``short rods'' region, \ie\ for fixed values of
the rod length $l$ that do not grow with $l_m$, $\beta l$ is a small
parameter in which we can expand. The limiting behaviour of the
product $\beta l_m$ is more difficult to determine, although it is
plausible that if this quantity is not large enough then the supposed
nematic shadow phase will behave as if $\beta=0$ exactly and thus
become fully identical to the isotropic cloud phase in the limit,
implying that no genuine phase coexistence has been found. We
therefore assume in the following that $\beta l_m \to\infty$ for
$l_m\to\infty$, and will find support for this in the numerical
results presented in Sec.~\ref{sec:numerical_support}.

Let us now turn to the behaviour of $\rho_2\N$ for
$l_m\to\infty$. Having shown that it cannot diverge, the two
possibilities are that $\rho_2\N\to$ const.\ $>0$ or $\rho_2\N\to
0$. In the first case we show in App.~\ref{app:rho2N=0} that the
problem can be drastically reduced to a single equation in
$\rho_2\N$. The only solution of this equation is $\rho_2\N=0$, in
contradicition with the assumed finite limit value. We therefore
conclude that both $\beta$ and $\rho_2\N $ vanish in the limit of
infinite cutoff. As explained above, we will see that the nematic
shadow phase remains a bona fide nematic in spite of the fact that
$\rho_2\N\to 0$: the longest rods exhibit strong nematic order but
their contribution to $\rho_2\N$ vanishes in the limit. As pointed out
earlier, the fact that
$\rho_2\N\to 0$ also demonstrates that the short rods in the nematic
shadow phase are orientationally disordered.

%%%%%%%%%%%%%%%%THE SHORT RODS BIT%%%%%%%%%%%%%%%%%%%%%%%%%%%%%% 
\subsection{Setup of the equations}\label{sec:theory}

We now proceed to simplify the equation for the isotropic cloud and
nematic shadow phases, using that $\beta$ and $\rho_2\N$ both converge
to zero for large $l_m$ while $\beta l_m$ diverges. The equations to
be solved are\eqqq{eq:rho0N}{eq:rho1N}{eq:rho2N} for the moments of the
nematic shadow phase, the osmotic pressure
equality~\eqref{eq:osmotic_pressure} and the
definition~\eqref{eq:beta} of $\beta$. 
These five equations determine the five unknowns $\rho_0$, $\rho_0\N$,
$\rho_1\N$, $\rho_2\N$ and $\beta$ as a function of the cutoff
$l_m$. Equivalently, we can use Eqs.\eqq{eq:rho0N}{eq:rho1N} to
eliminate $\rho_0\N$ and $\rho_1\N$, solving the three remaining
equations for $\rho_0$, $\rho_2\N$ and $\beta$.

We now proceed to implement our scheme of breaking the integrals over
$l$ into two intervals, namely $l=0\ldots\tilde l$ and $l=\tilde
l\ldots l_m$, where $1\ll\tilde l\ll l_m$. Calling $I_0$ the
contribution of the long rods to $\rho_0\N$, Eq.~\eqref{eq:rho0N}
becomes in this way
\[
\rho_0\N=\rho_0\int_0^{\tilde l} dl\ \normparent(l) e^{\beta l}\angint
e^{lf(\theta)}+I_0
\]
where 
\beq
\label{eq:I_0}
I_0=\rho_0 \int_{\tilde l}^{l_m} dl\ \normparent(l)e^{\beta l}\angint
e^{lf(\theta)}
\eeq
In the integral over the short rods we can expand in the small
parameters $\beta l$ and $\rho_2\N l$.  Since from Eq.~\eqref{eq:f}
$f(\theta)l$ is proportional to $\rho_2\N l$, an expansion to linear
order gives
\[
\rho_0\N = \rho_0\int_0^{\tilde l} dl\
\normparent(l)\angint\left\{1+\beta l+f(\theta)l\right\} + I_0
\]
Having isolated the contribution from the long rods, we can now extend
the integration to $\infty$ in the convergent $l$-integral in the
first term; the error from this approximation turns out to be
negligible compared to the cutoff-dependent terms that we focus on.
This gives, using that the average parental rod length is unity,
\beq\label{eq:rho0N_expanded}
\rho_0\N=\rho_0+(\beta-c_2\rho_2\N)\rho_0+I_0
\eeq
The same procedure can be applied to Eqs.\eqq{eq:rho1N}{eq:rho2N} and
yields
\begin{eqnarray}
\rho_1\N&=&\rho_0+(\beta-c_2\rho_2\N)\langle l^2\rangle\rho_0+I_1
\label{eq:rho1N_expanded}\\
\rho_2\N&=&\frac{c_2}{5}\rho_0\langle l^2\rangle\rho_2\N+I_2
\label{eq:rho2N_expanded1}
\end{eqnarray}
where the long rod parts are
\begin{eqnarray}
%\label{eq:long_rods_moments}
%I_0&=&\totint\parent(l)e^{\beta l}e^{lf(\theta)}\label{eq:I_0}\\
I_1&=& \rho_0
\int_{\tilde l}^{l_m} dl\ l\normparent(l)
e^{\beta l}e^{lf(\theta)}\label{eq:I_1}\\
I_2&=& \rho_0\int_{\tilde l}^{l_m} dl\ l\normparent(l)
e^{\beta l}\Ptwo e^{lf(\theta)}\label{eq:I_2}
\end{eqnarray}
and the averages $\langle l^n\rangle$ are taken over the normalized
parent distribution
\[
\langle l^n\rangle=\int dl\ l^n\normparent(l)
\]
The long rod contributions $I_0$, $I_1$, $I_2$ remain undetermined so
far. They must certainly be finite for $l_m\to\infty$, because the
moments are, but in fact will turn out to converge to zero. For $I_2$
this is easy to see from Eq.~\eqref{eq:rho2N_expanded1}, which implies
that $I_2\leq \rho_2\N$. We can further exclude the possibility that
$I_2\ll \rho_2\N$, which would imply from
Eq.~\eqref{eq:rho2N_expanded1} that
\[
\langle l^2\rangle\rho_0=4
\]
It is easy to show that this is just the spinodal condition for the
isotropic phase in the $\Ptwo$ Onsager model, while the onset of
isotropic-nematic phase coexistence must happen at a density below the
spinodal instability. Thus, $I_2$ must be of the same order as
$\rho_2\N$, which
means that short and long rods make comparable contributions to
$\rho_2\N$ as anticipated above.

To evaluate the integrals for $I_0$, $I_1$ and $I_2$ more
quantitatively, we note that they all contain the exponentially
diverging factor $\exp(\beta l)$ which dominates the behaviour of the
integrands since $\beta l_m\to\infty$. To leading order we can thus
approximate the integrands $\normparent(l)w(l)e^{\beta l}$, where
$w(l)$ is one of the weight functions in the above
Eqs.\eqqq{eq:I_0}{eq:I_1}{eq:I_2}, by $\normparent(l_m)w(l_m)e^{\beta
l}$ and integrate.  For the log-normal parent distribution we show in
App.~\ref{app:long_rods_integrals} that this approximation indeed
becomes exact for $l_m\to\infty$. For more quickly decaying
distributions, for example $\normparent(l)\sim \exp(-l^\alpha)$ with
$0<\alpha<1$, a slightly better approximation is needed (see
App.~\ref{app:long_rods_integrals}), but the integrals remain
dominated by their values at $l=l_m$. 

It follows from the above arguments that $I_0$ is smaller than $I_1$
by a factor of $1/l_m$ for large $l_m$, and can therefore be neglected
to leading order. Likewise, the ratio of $I_2$ and $I_1$ becomes
simply
\beq\label{eq:s_definition}
s=\frac{I_2}{I_1}
=\frac{\angint\Ptwo e^{l_mf(\theta)}}{\angint e^{l_mf(\theta)}}
\eeq
which is in fact identical to $S(l_m)$, the orientational order
parameter for the longest rods, and therefore lies between 0 and
1. Using the definition of $s$, Eq.~\eqref{eq:rho2N_expanded1} can be
written as 
\beq\label{eq:rho2N_expanded}
\rho_2\N =\frac{c_2}{5}\rho_0\langle l^2\rangle\rho_2\N +s\Ilm
\eeq
which will be useful shortly.

So far we have only dealt with the equations for the moment densities
$\rho_0\N$, $\rho_1\N$ and $\rho_2\N$ of the nematic phase. We now add
Eq.~\eqref{eq:beta} for $\beta$; inserting
Eq.~\eqref{eq:rho1N_expanded} and bringing all terms in $\beta$ to the
left gives
\beq\label{eq:beta_2}
\beta \left(1+c_1\langle l^2\rangle\rho_0\right)=
c_2\rho_2\N \left(1+c_1\langle l^2\rangle\rho_0\right)-c_1\Ilm
\eeq
This implies $\beta\leq c_2\rho_2\N$, and since $\beta l_m\to\infty$
for $l_m\to\infty$ we also have $\rho_2\N l_m \to\infty$. Recalling
$f(\theta)=c_2\rho_2\N \left(\Ptwo-1\right)$, we therefore see
from~\eqref{eq:s_definition} that the long rods have strong
orientational order. More quantitatively, by expanding
$\Ptwo-1\simeq -\frac{3}{2}\theta^2$ and performing the resulting
Gaussian integrals, 
%@@ certainly in the thesis you need to mention that
%you've restricted the integration range now to $\theta=0\ldots \pi/2$,
%otherwise you'd have to expand similarly around $\theta=\pi$ @@ 
we find that
\beq\label{eq:rhs_integral}
%\frac{\angint\left[\beta+c_2\rho_2\N \left(\Ptwo-1\right)\right]e^{l_mf(\theta)}}{\angint
%e^{l_mf(\theta)}}\simeq\frac{1}{3c_2\rho_2\N l_m}\left(\beta-\frac{1}{l_m}\right)
%3c_2\rho_2\N l_m
s=1-\frac{1}{c_2\rho_2\N l_m}
\eeq
to first order in $1/(\rho_2\N l_m)$. 
To leading order, one has $s=1$ and Eq.~\eqref{eq:rho2N_expanded}
simplifies to
\beq\label{eq:Ilm}
\Ilm=C\rho_2\N 
\eeq
with
\beq\label{eq:C_first}
C=1-\frac{c_2}{5}\rho_0\langle l^2\rangle
\eeq
Using this in Eq.~\eqref{eq:beta_2} we get:
\beq\label{eq:beta_linear}
\beta=\gamma\rho_2\N 
\eeq
where
\beq\label{eq:gamma}
\gamma=\frac{(6/5)c_1c_2\rho_0\langle l^2\rangle+c_2-c_1}{1+c_1\rho_0\langle
l^2\rangle}
\eeq
For our set of variables $\rho_0$, $\rho_2\N $ and $\beta$ we now have
a set of three equations, consisting of the simplified
Eqs.~\eqref{eq:Ilm} and~\eqref{eq:beta_linear} and the osmotic
pressure equality~\eqref{eq:osmotic_pressure}. Before turning to the
latter, we note that Eq.~\eqref{eq:gamma} implies that
$\beta$ is of the same order as $\rho_2\N$
for $l_m\to\infty$ unless $\gamma\rightarrow 0$ in the same limit. We
will find below that the second alternative holds, and that $\beta$
is actually proportional to $\left(\rho_2\N\right)^2$.

%\subsection{The pressure equality}

Consider now the pressure equality, given by
Eq.~\eqref{eq:osmotic_pressure} in its still exact form. We would like
to apply to this the same procedure of splitting the $l$-integrals
into contributions from short and long rods, and then expanding in
$\beta l$ and $\rho_2\N l$ in the short rods part. It turns out,
however, that the expansion in the short rods part needs to be
carried to {\em third} order in order to give an equation which is
independent of the ones we have already derived. We therefore first
rearrange the pressure equality into a form where the linear and
quadratic terms in the expansion cancel automatically.

On the l.h.s.\ of Eq.~\eqref{eq:osmotic_pressure}, we write $\rho_0 =
\lint \parent(l)$; in the quadratic term we similarly replace one of
the factors using $\rho_0=\rho_1=\lint \parent(l)l$. The r.h.s.\ can
be rewritten in a similar way, using the
expressions\eqqq{eq:rho0N}{eq:rho1N}{eq:rho2N} for the moments of the
nematic phase. Including an overall factor of 2, the osmotic pressure
equality then takes the form
\begin{eqnarray}
0&=&\totint\parent(l)\left\{e^{\beta
l+f(\theta)l}\left[2+c_1\rho_1\N l - c_2\rho_2\N l \Ptwo \right]\right.
\nonumber\\
& & -\left. 2-c_1\rho_0 l \right\}
\label{eq:PI_new}
\end{eqnarray}
and the definitions~\eqref{eq:f} and~\eqref{eq:beta} can be used to
write the last two terms in the square bracket as
\beq
\label{eq:rewrite}
c_1\rho_1\N l - c_2\rho_2\N l \Ptwo
= [-f(\theta)-\beta+c_1\rho_0]l
\eeq
In the same way, the definition~\eqref{eq:beta_def} of $\beta$ can be
written in integral form as
\[
\beta=\totint
\parent(l)c_1 l\left[1-e^{\beta l+f(\theta)l}\right]+c_2\rho_2\N 
\]
Multiplying by $\rho_0=\lint \parent(l)l$ and bringing all terms to
the r.h.s.\ gives
\beq\label{eq:beta_new}
0= \totint\parent(l)\left\{-\beta l+c_1\rho_0
l\left[1-e^{\beta
l+f(\theta)l}\right]+c_2\rho_2\N l\right\}
\eeq
If we now sum Eq.~\eqref{eq:PI_new} and Eq.~\eqref{eq:beta_new}, and
make use of Eq.~\eqref{eq:rewrite} and the fact that $c_2\rho_2\N =
-\angint f(\theta)$, we obtain after a few rearrangements
\begin{eqnarray}
0&=&\totint\parent(l)\left\{2\left[e^{\beta
l+f(\theta)l}-1\right]\right.
\nonumber\\
& &-\left.\left[\beta+f(\theta)\right]l\left[e^{\beta
l+f(\theta)l}+1\right]\right\}
\label{eq:pressure_transformed}
\end{eqnarray}
Eq.~\eqref{eq:pressure_transformed} is the desired form of the osmotic
pressure equality, which is still exact. We can now proceed as above
and split the $l$-integral into the contributions from short and long
rods. In the short rods part, we Taylor expand the integrand in the
small variable $[\beta+f(\theta)]l$; as anticipated above, the first
nonzero term in this expansion is of third order.
To leading order, Eq.~\eqref{eq:pressure_transformed} then becomes
\beq\label{eq:lhs_expanded}
0 = -\int_0^{\tilde l} dl\ \thint\
\parent(l)\frac{l^3}{6}\left[\beta+f(\theta)\right]^3 + I_3
\eeq
where $I_3$ represents the long rods part of the
integral in Eq.~\eqref{eq:pressure_transformed}. In the first term we
can again take the upper integration limit to $\infty$ and obtain
\beq\label{eq:lhs_final}
-\frac{\langle l^3\rangle}{6}\rho_0\angint\left[\beta+f(\theta)\right]^3=-I_3
\eeq
%
%where the average is over the normalized parent distribution
%$P^{(0)}(l)=\parent(l)/\parent_0$.
%
where as above the average $\langle l^3\rangle$ is over the normalized
parent distribution $\normparent(l)$.

Our final task in simplifying the pressure equality is to find $I_3$.
Because $\beta l_m$ diverges (and we have restricted the
integration range to the long rods) the exponential factors in
Eq.~\eqref{eq:pressure_transformed} dominate the integral. Of the
terms involving these factors, the first one is also negligible compared 
to the second one, $-[\beta +\ft] l e^{\beta l+\ft l}$, since it is
missing the large factor $[\beta + \ft]l$. Comparing with
Eqs.\eqq{eq:I_1}{eq:I_2}, we thus have that 
\[
I_3=-[\beta I_1+c_2\rho_2\N(I_2-I_1)] = -[\beta+c_2\rho_2\N(s-1)]\Ilm
\]
for large $l_m$. Eq.~\eqref{eq:rhs_integral} further implies that
$c_2\rho_2\N(s-1)=-1/l_m$ to leading order, which is negligible
compared to $\beta$ since $\beta l_m\to\infty$. Altogether we have the
simple result $I_3=-\beta \Ilm$, and the transformed pressure
equality~\eqref{eq:pressure_transformed} becomes in its final form
\beq
\label{eq:expanded_final}
\label{eq:pressure_expanded}
-\frac{\langle l^3\rangle}{6}\rho_0\angint\left[\beta+c_2\rho_2\N
\left(\Ptwo-1\right)\right]^3=\beta \Ilm
\eeq

%%%%%%%%%%%%%%%%%%%%%%%SOLUTION%%%%%%%%%%%%%%%%%%%%%%%%%%%%%%%%%%%%%%
\subsection{The isotropic cloud point} \label{sec:CPC}

In the previous subsection, we used the large cutoff-limit to simplify
the three basic equations for the three variables $\rho_0$, $\rho_2\N$
and $\beta$, obtaining Eq.~\eqref{eq:Ilm} for $\rho_2$,
Eq.~\eqref{eq:beta_linear} for $\beta$ and the transformed pressure
equality~\eqref{eq:expanded_final} for $\rho_0$. While in principle
these equations need to be solved simultaneously, the limiting value
of the isotropic cloud point is easy to extract directly, as follows.
Using Eqs.\eqq{eq:Ilm}{eq:beta_linear} to eliminate $\beta$ and
$I_1$, and dividing by $\left(\rho_2\N\right)^2$ the pressure
equality~\eqref{eq:expanded_final} becomes
\beq\label{eq:pressure_contradiction}
-\frac{\langle l^3\rangle}{6}\rho_0\rho_2\N \angint\left[\gamma+c_2\left(\Ptwo-1\right)\right]^3=C\gamma
\eeq
%
%\beq\label{eq:rhs_final}
%-\frac{\langle l^3\rangle}{6}\rho_0\angint\left[\beta+c_2\rho_2\N \left(\Ptwo-1\right)\right]^3=C\beta\rho_2\N 
%\eeq
%On the lhs instead, using Eq.~\eqref{eq:beta_linear}, we have
%\beq
%-\frac{\phi^{(0)}_3}{6}\left(\rho_2\N \right)^3\angint\left[\gamma+c_2\left(\Ptwo-1\right)\right]^3
%\eeq
%So rearranging, the pressure equality becomes:
%
The l.h.s.\ of this equation tends to zero for $l_m\to\infty$ since
$\rho_2\N$ does; in the same limit we must therefore have $C\gamma=0$.
Thus either $C=0$ or $\gamma=0$; but the first case we have already
excluded in our discussion after Eq.~\eqref{eq:rho2N_expanded1} as
leading to the spinodal instability rather than the cloud point.
Thus $\gamma=0$, and inserting $c_1=2$ and $c_2=5/4$ into
Eq.~\eqref{eq:gamma} gives
\beq\label{eq:CPC_rho_0}
\langle l^2\rangle\rho_0=\frac{1}{4}
\eeq
as our final result for the parent density $\rho_0$ at the isotropic
cloud point, in the large cutoff limit $l_m\to\infty$. From the
definition~\eqref{eq:sigma} of the polydispersity $\sigma$, and
bearing in mind that we took the average rod length in the parent to
be unity, we have $\langle l^2\rangle=\sigma^2+1$ so that the cloud
point condition can also be written as
\[
\rho_0=\frac{1}{4}\frac{1}{\sigma^2+1}
\]
The result $\gamma\rightarrow 0$ for $l_m\to \infty$ implies, from
Eq.~\eqref{eq:beta_linear}, that $\beta\ll\rho_2\N$ for large $l_m$.
This relation helps to understand the appearance of the minus sign on
the l.h.s.\ of the transformed pressure
equality~\eqref{eq:pressure_expanded}. Call
$g(\theta)=\beta+c_2\rho_2\N (\Ptwo-1)$ the function appearing in the
integrand. Then $\beta=g(0)>0$ while for other $\theta$ the value of
$g(\theta)$ is lower. In fact, if $\beta\ll\rho_2\N$, then $g(\theta)$
is {\em negative} except in a very small region around $\theta=0$ (and
of course $\theta=\pi$). These negative values of $g(\theta)$ will
dominate the integral on the l.h.s.\ of
Eq.~\eqref{eq:pressure_expanded} and cause it to be negative, making
the l.h.s.\ positive overall.  Interestingly, the situation on the
r.h.s.\ is the reverse. As discussed above, for large $l_m$ we can
write
\begin{eqnarray*}
-I_3 &=& \totint \parent(l) [\beta +\ft] l e^{\beta l+\ft l} \\
&=& \totint \parent(l) g(\theta)l e^{g(\theta)l}
\end{eqnarray*}
Since the integral here is over the long rods only, and in addition
$g(0)l_m = \beta l_m \to\infty$ for $l_m\to\infty$, the exponential
factor ensures that the integral is dominated entirely by the
small region of angles around $\theta=0$ (and $\theta=\pi$) where
$g(\theta)$ is positive.

Having seen that $\beta\ll\rho_2\N$ for large $l_m$, we can now show
that in fact $\beta\sim\left(\rho_2\N \right)^2$.  On the l.h.s.\ of
Eq.~\eqref{eq:expanded_final} we can neglect the first term in square
brackets and integrate the remainder explicitly to obtain
$(54/35)({\langle l^3\rangle}/{6})\rho_0c_2^3\left(\rho_2\N
\right)^3$.  On the r.h.s.\ nothing changes and we can use
Eq.~\eqref{eq:Ilm} to eliminate $\Ilm$. Dividing by $\rho_2\N$,
Eq.~\eqref{eq:expanded_final} thus becomes
\beq\label{eq:pressure_for_beta}
\beta=\kappa\left(\rho_2\N \right)^2, \qquad
\kappa = \frac{9}{35}\frac{\langle l^3\rangle\rho_0c_2^3}{C}
\eeq
showing that indeed $\beta\propto\left(\rho_2\N \right)^2$.

%%%%%%%%%%%%%%%%%%%%%%%%%%%%%%%%%Beta scaling%%%%%%%%%%%%%%%%%%
\subsection{Scaling of $\beta$ and $\rho_2\N $}\label{sec:scaling}

So far we have only used our simplified equations for the isotropic
cloud point to deduce the value of the cloud point density for
$l_m\to\infty$. The equations were obtained by expanding the short rod
integrals in the equations for $\rho_2\N$ and $\beta$ to linear order
in these two variables; in the osmotic pressure equality we needed to
expand to third order because lower terms cancel out automatically.
From the expanded pressure equality we deduced that the 
coefficient $\gamma$ relating $\beta$ and $\rho_2\N$ to linear order 
has to vanish in the limit $l_m\rightarrow\infty$, and that in fact
$\beta$ must be proportional to $(\rho_2\N)^2$ to leading order; the
condition $\gamma=0$ gave us the limiting value of the cloud point
density for infinite cutoff. These results all apply generally for
parent rod length distributions with fat (less than exponentially
decaying) tails.

We now proceed to obtain from our equations additional information
regarding the variation of the key quantities with the cutoff $l_m$,
in the regime of large $l_m$. These quantities are $\beta$, $\rho_2\N$
and the moment densities of the isotropic cloud and nematic shadow
phases. To do so, we have to explicitly calculate the long rod
contribution
\[
\Ilm= \rho_0 \int_{\tilde l}^{l_m} dl\ \normparent(l)l e^{\beta
l}\angint e^{lf(\theta)} 
\]
to $\rho_1\N$, which so far we had left unevaluated.  Anticipating
that the integral is dominated by rod lengths of order $l_m$, for
which $\rho_2\N l$ is large since $\rho_2\N l_m\to\infty$ as
$l_m\to\infty$, we can evaluate the angular integral by expanding
around $\theta=0$ (see before Eq.~\eqref{eq:rhs_integral}) to get
\beq\label{eq:nematic_moments_sno}
\Ilm=\rho_0 \int_{\tilde l}^{l_m}dl\ 
\normparent(l) \frac{e^{\beta l}}{3c_2\rho_2\N}
\eeq
It will be useful to remove the overall factors and define the quantity
\beq\label{eq:II}
\II_1 = \frac{3c_2\rho_2\N}{\rho_0}\Ilm = \int_{\tilde l}^{l_m}dl\ 
\normparent(l) e^{\beta l}
\eeq
which only depends on $\beta$ and $l_m$. From Eq.~\eqref{eq:Ilm} we
then have $\II_1 = (3c_2 C/\rho_0)\left(\rho_2\N\right)^2$, and using
Eq.~\eqref{eq:pressure_for_beta} to eliminate $\rho_2\N$ in favour of
$\beta$ gives
\beq
\label{eq:beta_lm}
\frac{\II_1}{\beta} = A, \qquad A=\frac{3c_2 C}{\rho_0\kappa}
\eeq
All factors in the definition of $A$, and therefore $A$ itself, tend
to nonzero constants as $l_m\to\infty$; the above equation therefore
implicitly determines the asymptotic dependence of $\beta$ on $l_m$.

To evaluate $\II_1$, one notes (see App.~\ref{app:long_rods_integrals})
that for fat-tailed parent distributions such as the log-normal the
exponentially increasing factor $\exp(\beta l)$ dominates the
behaviour of the integrand. The integral is therefore dominated by the
region $l\approx l_m$ and one can replace $l=l_m$ in the
non-exponential factors of the integrand, yielding
\beq
\label{eq:Ilm_peak}
\label{eq:asymptotic_Ilm}
\II_1=\normparent(l_m)\frac{e^{\beta l_m}}{\beta}
\eeq
for large $l_m$. Eq.~\eqref{eq:beta_lm} then gives
\beq\label{eq:beta*l_m}
\frac{e^{\beta l_m}}{(\beta l_m)^2}=\frac{A}{l_m^2 \normparent(l_m)}
\eeq
The function $e^x/x^2$ on the lhs of Eq.~\eqref{eq:beta*l_m} can
be inverted for large argument $x$ and yields asymptotically just a
logarithm (see App.~\ref{app:long_rods_integrals}), so that
\beq\label{eq:beta_scaling_parent}
\beta=\frac{1}{l_m}\ln\left(
\frac{A}{l_m^2 \normparent(l_m)}
\right)
\eeq
Now for a log-normal parent distribution $\ln\normparent(l) =
-\ln^2l_m/(2 w^2)$ plus terms that are negligible by comparison
for large $l_m$, giving for the leading asymptotic dependence of $\beta$ on
$l_m$
\beq\label{eq:beta_scaling_lognormal}
\beta = \frac{\ln^2 l_m}{2l_m w^2}
\eeq
For a more general parent distribution with a fat tail, \ie\
$\normparent(l)=e^{-h(l)}$ with $h(l)$ a function that diverges with
$l$ less than linearly, the approximation~\eqref{eq:asymptotic_Ilm} is
not quite correct but can be refined (see
App.~\ref{app:long_rods_integrals}) to
\beq
\II_1 = \frac{e^{\beta l_m - h(l_m)}}
{\beta - h^\prime(l_m)}
\label{eq:II_refined}
\eeq
Inserting into Eq.~\eqref{eq:beta_lm}, one can again solve for $\beta$
and obtains asymptotically
\beq\label{eq:beta_scaling_general}
\beta = \frac{h(l_m)}{l_m}
\eeq
which generalizes Eq.~\eqref{eq:beta_scaling_lognormal}. For $h(l)
\sim l^\alpha$ with $0<\alpha<1$, which corresponds to a parent
distribution decaying more quickly than a log-normal but still less
than exponentially, one gets $\beta\sim l_m^{\alpha-1}$. Using that
$\beta \propto \left(\rho_2\N\right)^2$ for large $l_m$, the above
results also imply that $\rho_2\N\sim [h(l_m)/l_m]^{1/2}$, with
$\rho_2\N \propto l_m^{-1/2}\ln l_m$ for the log-normal case.  Notice
that the results for $h(l)\sim l^\alpha$ extrapolate sensibly towards
the case of a parent distribution without a fat tail, \ie\
$\alpha=1$. Then neither $\beta$ nor $\rho_2\N$ have any reason to
tend to zero for $l_m\to\infty$, and this is consistent with the limit
$\alpha\to 1$ of the scalings above. As explained in
Sec.~\ref{sec:cloud_shadow}, the scaling of $\rho_2\N$ also determines
directly the length-dependence of orientational order in the nematic
phase: only rods with $l$ around the ``ordering length'' $1/\rho_2\N$
and greater show significant ordering. For the log-normal case, the
ordering length is $1/\rho_2\N \sim l_m^{1/2}/\ln l_m$ and $\gg 1$ as
expected, but simultaneously $\ll l_m$. This makes precise our earlier
statements that orientational order is confined to the longer rods in
the nematic phase; ordering appears for rod lengths far above unity,
but still well below the cutoff length $l_m$.

\subsection{Scaling of densities} \label{sec:density_scaling}

Finally, we study the scaling with $l_m$ of the isotropic cloud point density
$\rho_0$ and the number density $\rho_0\N$ and rescaled volume
fraction $\rho_1\N$ of the nematic shadow phase.
From Eqs.\eqq{eq:rho0N_expanded}{eq:rho1N_expanded}, with
$\beta\ll\rho\N_2$ and $I_0\ll I_1$ and inserting
Eq.~\eqref{eq:Ilm}, one gets
\begin{eqnarray}
\rho_0\N &=&\rho_0-c_2\rho_0\rho_2\N \label{eq:scaling_rho0N_rho2}\\ 
\rho_1\N &=&\rho_0+\left[C-c_2\rho_0\langle l^2\rangle
\right]\rho_2\N 
\label{eq:scaling_rho1N_rho2}
\end{eqnarray}
where $\rho_0$ is the density of the isotropic cloud phase at finite,
large cutoff, which is still a function of $l_m$.  The interpretation
of these results is that for finite but large $l_m$, the cloud and
shadow curves differ by terms of order $\rho_2\N$ ($ \sim
l_m^{-1/2}\ln l_m$ for a log-normal parent), whether expressed in
terms of $\rho_0$ or $\rho_1$. In the limit $l_m\rightarrow\infty$,
the shadow curve collapses onto the cloud curve, and both tend
asymptotically to the limiting form~\eqref{eq:CPC_rho_0}.
This may appear counterintuitive, suggesting that the nematic phase
does not at all ``feel'' the presence of the orientational order. The
answer is of course that only the longer rods in the nematic are
orientationally ordered, and for rod lengths comparable to the cutoff
$l_m$ the nematic and isotropic density distributions do indeed
differ. That there is a genuine difference between the isotropic cloud and
nematic shadow phases becomes evident if we look at higher order
moment densities $\lint \rho(l)l^n$ of the density distributions. In
the parent, and therefore the isotropic cloud phase, all these moments
will generically be finite (one exception being parent distributions
with power-law tails). For the nematic phase, the picture is
completely different. For $n=1$ we have the moment $\rho_1\N$, and we
showed above that the long rod contribution to this converges to zero
for $l_m\to\infty$. For larger $n>1$, since the long rod integrals are
dominated by values of $l\simeq l_m$, the long rod contribution will
scale as $l_m^{n-1}I_1\sim l_m^{n-1}\rho_2\N$ from Eq.~\eqref{eq:Ilm}
(see App.~\ref{app:long_rods_integrals}). For the log-normal parent,
we found that $\rho_2\N \sim l_m^{-1/2} \ln l_m$ and so all moments of
the nematic shadow phase of order $n>3/2$ {\em diverge} with
$l_m$. For parents with less fat tails, $\normparent \sim
\exp(-l^{\alpha})$, we had instead $\beta\sim l_m^{\alpha-1}$ and
$\rho_2\N\sim \beta^{1/2} \sim l_m^{(\alpha-1)/2}$, implying that
nematic moments of order $n>(3-\alpha)/2$ diverge. Similarly, moment
densities taking account of orientational order, $\totint
\rho\N(l,\theta)l^n \Ptwo$ diverge when the exponent $n$ is large
enough. As we can see then, the similarities between the isotropic cloud phase
and the nematic shadow phase are rather more superficial than
suggested by the fact that their low-order ``physical'' moment
densities $\rho_0$ and $\rho_1$ agree, and that $\rho_2\N\to 0$ for
$l_m\to\infty$; with respect to any quantity that attaches sufficient
weight to the behaviour of the long rods, they are extremely
different.

To complete our analysis of the scaling of the moment densities, we
need to know in
Eqs.\eqq{eq:scaling_rho0N_rho2}{eq:scaling_rho1N_rho2} how the cloud
point density $\rho_0$ depends on $l_m$. By Taylor expanding the
relation between $\gamma$ and $\rho_0$, Eq.~\eqref{eq:gamma}, one
easily sees that the deviation of $\rho_0$ from its limiting value
$1/(4\langle l^2\rangle)$ is, to lowest order, proportional to
$\gamma$. To determine how $\gamma$ converges to zero, compare
$\beta=\kappa\left(\rho_2\N \right)^2$ from
Eq.~\eqref{eq:pressure_for_beta} with $\beta=\gamma\rho_2\N$ from
Eq.~\eqref{eq:beta_linear}. This shows that $\gamma$ should be
proportional to $\rho_2\N$, although we cannot determine the
coefficient explicitly: the result $\beta=\gamma\rho_2\N$ is only the
lowest order in an expansion in $\rho_2\N$ and we would need all terms
contributing at quadratic order to get the proportionality coefficient
in $\gamma \propto \rho_2\N$. Nevertheless, $\gamma\propto \rho_2\N$
alone tells us that $\rho_0$ approaches its limiting value as
\beq\label{eq:scaling_rhoI_rho2}
\rho_0=\frac{1}{4}\frac{1}{\sigma^2+1}+{\mathcal{O}}(\gamma)
=\frac{1}{4}\frac{1}{\sigma^2+1}+{\mathcal{O}}(\rho_2\N)
\eeq
This result, together with
Eqs.\eqq{eq:scaling_rho0N_rho2}{eq:scaling_rho1N_rho2} shows that for
large $l_m$ all the densities $\rho_0$, $\rho_0\N$ and $\rho_1\N$
differ from their limiting value for $l_m\to\infty$ by terms
proportional to $\rho_2\N$.

\subsection{Numerical support for the theory}
\label{sec:numerical_support}

The analytical results obtained so far are based on the fact that
$\beta$ and $\rho_2\N$ must converge to zero as the cutoff $l_m$
becomes large; we supplemented this with the assumption that $\beta
l_m$ diverges, from which it then followed that also $\rho_2\N l_m$
diverges. These statements allowed us to argue that integrals over the
rod length $l$ could be split into two distinct parts: one for the
short rods, where we could expand in the small parameters $\beta$ and
$\rho_2\N$, and one for the long rods where the exponential factor
$\exp(\beta l)$ is large and dominant and leads to an increase in the
nematic density distribution towards the cutoff length. We already saw
some graphical support for this in Fig.~\ref{fig:I_N_distr_both2} but
provide in this section more detailed numerical support for the
theory, focussing on the case of a log-normal parent distribution.

From the theory we obtained that in the limit of infinite cutoff, the
cloud and shadow curves collapse onto the curve given by
Eq.~\eqref{eq:CPC_rho_0}. In other words, the moment densities
$\rho_0=\rho_1$ of the isotropic cloud phase, and those of the nematic
phase ($\rho_0\N$ and $\rho_1\N$) should tend to the same limit for
$l_m\rightarrow\infty$. A plot of these densities against $l_m$
(Fig.~\ref{fig:asy_moms}), for fixed polydispersity $\sigma$, shows
results compatible with this prediction, with all curves tending
towards the theoretically predicted limiting value. A more detailed
check is obtained by plotting the densities against $\rho_2\N$. Our
theory predicts that they should approach their common asymptotic
value linearly in $\rho_2\N$, and Fig.~\ref{fig:moms_scale_rho2} (a)
confirms this. We can even check the known coefficients of this linear
dependence by plotting the differences $\rho_0-\rho_0\N$ and
$\rho_1\N-\rho_0$ against $\rho_2\N$.
Eqs.\eqq{eq:scaling_rho0N_rho2}{eq:scaling_rho1N_rho2} predict that
$\rho_0-\rho_0\N=c_2\rho_0\rho_2\N$ and
$\rho_1\N-\rho_0=[C-c_2\rho_0\langle l^2\rangle]\rho_2\N$. In this
representation, the $l_m\to\infty$ limits of the coefficients of
$\rho_2\N$ on the right-hand sides are known, and 
the unknown proportionality factor in the approach of $\rho_0$ to its
limiting value (see Eq.~\eqref{eq:scaling_rhoI_rho2}) has been eliminated.
Figure~\ref{fig:moms_scale_rho2} (b) displays the theoretically
predicted linear relations and shows good convergence of the numerical
results to these asymptotes.
\bfig
\begin{picture}(0,0)%
\includegraphics{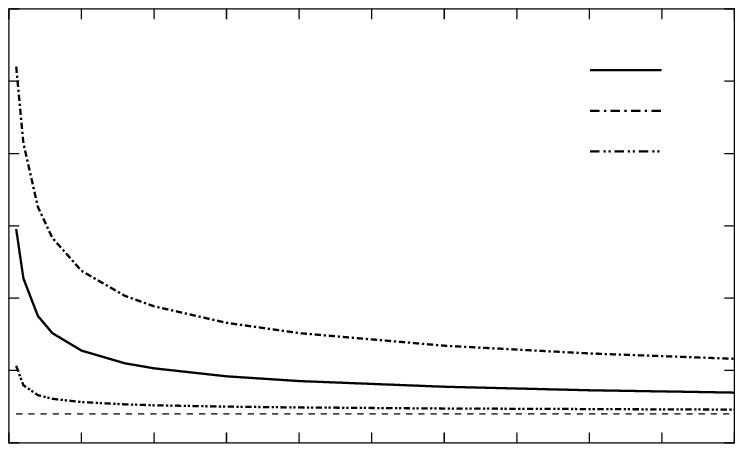}%
\end{picture}%
\setlength{\unitlength}{2565sp}%
\begingroup\makeatletter\ifx\SetFigFont\undefined
% extract first six characters in \fmtname
\def\x#1#2#3#4#5#6#7\relax{\def\x{#1#2#3#4#5#6}}%
\expandafter\x\fmtname xxxxxx\relax \def\y{splain}%
\ifx\x\y   % LaTeX or SliTeX?
\gdef\SetFigFont#1#2#3{%
  \ifnum #1<17\tiny\else \ifnum #1<20\small\else
  \ifnum #1<24\normalsize\else \ifnum #1<29\large\else
  \ifnum #1<34\Large\else \ifnum #1<41\LARGE\else
     \huge\fi\fi\fi\fi\fi\fi
  \csname #3\endcsname}%
\else
\gdef\SetFigFont#1#2#3{\begingroup
  \count@#1\relax \ifnum 25<\count@\count@25\fi
  \def\x{\endgroup\@setsize\SetFigFont{#2pt}}%
  \expandafter\x
    \csname \romannumeral\the\count@ pt\expandafter\endcsname
    \csname @\romannumeral\the\count@ pt\endcsname
  \csname #3\endcsname}%
\fi
\fi\endgroup
\begin{picture}(5454,3885)(659,-3419)
\put(3751,-3361){\makebox(0,0)[lb]{\smash{\SetFigFont{8}{9.6}{rm}{\color[rgb]{0,0,0}$l_m$}%
}}}
\put(4501,-436){\makebox(0,0)[lb]{\smash{\SetFigFont{8}{9.6}{rm}{\color[rgb]{0,0,0}$\rho_1\N$}%
}}}
\put(4501,-136){\makebox(0,0)[lb]{\smash{\SetFigFont{8}{9.6}{rm}{\color[rgb]{0,0,0}$\rho_0$}%
}}}
\put(4501,-736){\makebox(0,0)[lb]{\smash{\SetFigFont{8}{9.6}{rm}{\color[rgb]{0,0,0}$\rho_0\N$}%
}}}
\put(659,-2875){\makebox(0,0)[rb]{\smash{\SetFigFont{8}{9.6}{rm}{\color[rgb]{0,0,0}0}%
}}}
\put(659,-2341){\makebox(0,0)[rb]{\smash{\SetFigFont{8}{9.6}{rm}{\color[rgb]{0,0,0}0.5}%
}}}
\put(659,-1807){\makebox(0,0)[rb]{\smash{\SetFigFont{8}{9.6}{rm}{\color[rgb]{0,0,0}1}%
}}}
\put(659,-1273){\makebox(0,0)[rb]{\smash{\SetFigFont{8}{9.6}{rm}{\color[rgb]{0,0,0}1.5}%
}}}
\put(659,-739){\makebox(0,0)[rb]{\smash{\SetFigFont{8}{9.6}{rm}{\color[rgb]{0,0,0}2}%
}}}
\put(659,-205){\makebox(0,0)[rb]{\smash{\SetFigFont{8}{9.6}{rm}{\color[rgb]{0,0,0}2.5}%
}}}
\put(659,329){\makebox(0,0)[rb]{\smash{\SetFigFont{8}{9.6}{rm}{\color[rgb]{0,0,0}3}%
}}}
\put(733,-2999){\makebox(0,0)[b]{\smash{\SetFigFont{8}{9.6}{rm}{\color[rgb]{0,0,0}0}%
}}}
\put(1269,-2999){\makebox(0,0)[b]{\smash{\SetFigFont{8}{9.6}{rm}{\color[rgb]{0,0,0}500}%
}}}
\put(1805,-2999){\makebox(0,0)[b]{\smash{\SetFigFont{8}{9.6}{rm}{\color[rgb]{0,0,0}1000}%
}}}
\put(2340,-2999){\makebox(0,0)[b]{\smash{\SetFigFont{8}{9.6}{rm}{\color[rgb]{0,0,0}1500}%
}}}
\put(2876,-2999){\makebox(0,0)[b]{\smash{\SetFigFont{8}{9.6}{rm}{\color[rgb]{0,0,0}2000}%
}}}
\put(3412,-2999){\makebox(0,0)[b]{\smash{\SetFigFont{8}{9.6}{rm}{\color[rgb]{0,0,0}2500}%
}}}
\put(3948,-2999){\makebox(0,0)[b]{\smash{\SetFigFont{8}{9.6}{rm}{\color[rgb]{0,0,0}3000}%
}}}
\put(4484,-2999){\makebox(0,0)[b]{\smash{\SetFigFont{8}{9.6}{rm}{\color[rgb]{0,0,0}3500}%
}}}
\put(5019,-2999){\makebox(0,0)[b]{\smash{\SetFigFont{8}{9.6}{rm}{\color[rgb]{0,0,0}4000}%
}}}
\put(5555,-2999){\makebox(0,0)[b]{\smash{\SetFigFont{8}{9.6}{rm}{\color[rgb]{0,0,0}4500}%
}}}
\put(6091,-2999){\makebox(0,0)[b]{\smash{\SetFigFont{8}{9.6}{rm}{\color[rgb]{0,0,0}5000}%
}}}
\end{picture}
\vspace*{0.2cm}
\caption{Plot of the cloud point density $\rho_0$ and the nematic
shadow density $\rho_0\N$ and rescaled volume fraction $\rho_1\N$
against the cutoff $l_m$, for fixed polydispersity $\sigma\approx
0.532$ ($w=0.5$). The
horizontal line shows the common limiting value for $l_m\to\infty$
which our theory predicts, $1/[4(\sigma^2+1)]\approx 0.195$.}
\label{fig:asy_moms}
\efig
\bfig
\begin{picture}(0,0)%
\includegraphics{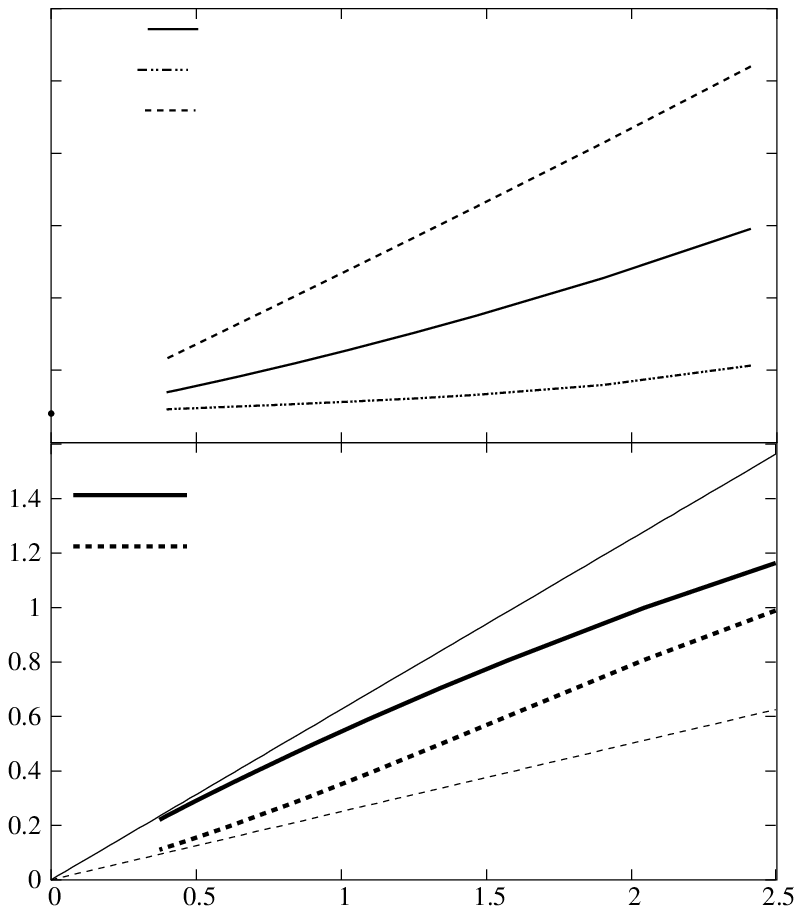}%
\end{picture}%
\setlength{\unitlength}{2565sp}%
\begingroup\makeatletter\ifx\SetFigFont\undefined
% extract first six characters in \fmtname
\def\x#1#2#3#4#5#6#7\relax{\def\x{#1#2#3#4#5#6}}%
\expandafter\x\fmtname xxxxxx\relax \def\y{splain}%
\ifx\x\y   % LaTeX or SliTeX?
\gdef\SetFigFont#1#2#3{%
  \ifnum #1<17\tiny\else \ifnum #1<20\small\else
  \ifnum #1<24\normalsize\else \ifnum #1<29\large\else
  \ifnum #1<34\Large\else \ifnum #1<41\LARGE\else
     \huge\fi\fi\fi\fi\fi\fi
  \csname #3\endcsname}%
\else
\gdef\SetFigFont#1#2#3{\begingroup
  \count@#1\relax \ifnum 25<\count@\count@25\fi
  \def\x{\endgroup\@setsize\SetFigFont{#2pt}}%
  \expandafter\x
    \csname \romannumeral\the\count@ pt\expandafter\endcsname
    \csname @\romannumeral\the\count@ pt\endcsname
  \csname #3\endcsname}%
\fi
\fi\endgroup
\begin{picture}(5906,7079)(336,-6613)
\put(976,164){\makebox(0,0)[lb]{\smash{\SetFigFont{8}{9.6}{rm}{\color[rgb]{0,0,0}$\rho_0$}%
}}}
\put(976,-436){\makebox(0,0)[lb]{\smash{\SetFigFont{8}{9.6}{rm}{\color[rgb]{0,0,0}$\rho_1\N$}%
}}}
\put(976,-136){\makebox(0,0)[lb]{\smash{\SetFigFont{8}{9.6}{rm}{\color[rgb]{0,0,0}$\rho_0\N$}%
}}}
\put(2815,-6563){\makebox(0,0)[lb]{\smash{\SetFigFont{8}{9.6}{rm}{\color[rgb]{0,0,0}$\rho_2\N$}%
}}}
\put(1886,-3653){\makebox(0,0)[lb]{\smash{\SetFigFont{8}{9.6}{rm}{\color[rgb]{0,0,0}$\rho\N_1-\rho_0$}%
}}}
\put(1886,-3277){\makebox(0,0)[lb]{\smash{\SetFigFont{8}{9.6}{rm}{\color[rgb]{0,0,0}$\rho_0-\rho\N_0$}%
}}}
\put(6151,-1261){\makebox(0,0)[lb]{\smash{\SetFigFont{8}{9.6}{rm}{\color[rgb]{0,0,0}(a)}%
}}}
\put(6151,-4486){\makebox(0,0)[lb]{\smash{\SetFigFont{8}{9.6}{rm}{\color[rgb]{0,0,0}(b)}%
}}}
\put(659,-2875){\makebox(0,0)[rb]{\smash{\SetFigFont{8}{9.6}{rm}{\color[rgb]{0,0,0}0}%
}}}
\put(659,-2341){\makebox(0,0)[rb]{\smash{\SetFigFont{8}{9.6}{rm}{\color[rgb]{0,0,0}0.5}%
}}}
\put(659,-1807){\makebox(0,0)[rb]{\smash{\SetFigFont{8}{9.6}{rm}{\color[rgb]{0,0,0}1}%
}}}
\put(659,-1273){\makebox(0,0)[rb]{\smash{\SetFigFont{8}{9.6}{rm}{\color[rgb]{0,0,0}1.5}%
}}}
\put(659,-739){\makebox(0,0)[rb]{\smash{\SetFigFont{8}{9.6}{rm}{\color[rgb]{0,0,0}2}%
}}}
\put(659,-205){\makebox(0,0)[rb]{\smash{\SetFigFont{8}{9.6}{rm}{\color[rgb]{0,0,0}2.5}%
}}}
\put(659,329){\makebox(0,0)[rb]{\smash{\SetFigFont{8}{9.6}{rm}{\color[rgb]{0,0,0}3}%
}}}
\end{picture}
\vspace*{0.2cm}
\caption{(a) Same as Fig.~\protect\ref{fig:asy_moms}, but with $\rho_2\N$
(which itself is a function of the cutoff $l_m$) shown on the $x$-axis
instead of $l_m$ itself. The results are compatible with the
theoretically predicted linear approach, for $\rho_2\N\to 0$, to the
common limiting value of $1/4(\sigma^2+1)\approx 0.195$
($\sigma\approx 0.532$) that is marked by the
dot. (b) Differences $\rho_0-\rho\N_0$ (bold solid) and
$\rho\N_1-\rho_0$ (bold dashed) versus $\rho\N_2$. For large $l_m$ (small
$\rho_2\N$), these converge to the linear relations predicted by
Eqs.~\eqref{eq:scaling_rho0N_rho2} and~\eqref{eq:scaling_rho1N_rho2}
and shown by the thin lines.
}
\label{fig:moms_scale_rho2}
\efig
From the theory we also obtained the scaling of $\beta$ with $l_m$,
giving $\beta = (\ln^2l_m)/(2l_m w^2)$ to leading order. In
Fig.~\ref{fig:beta_scale_lm} we check this by comparing plots of
$\beta$ and $(\ln^2l_m)/(2l_m w^2)$, both against $l_m$. For a
more detailed comparison, we can estimate from our theory the first
correction to the leading behaviour of $\beta$. From
Eq.~\eqref{eq:beta_scaling_parent} we see for a log-normal parent
distribution that this correction term, which arises both from the
factor $l_m^2$ in the denominator and the subleading term in
$\ln\normparent(l_m)$, should scale as $\sim l_m^{-1}\ln l_m$.
(Again, because we have not systematically kept track of subleading
terms, we do not attempt to determine the coefficient of this
correction term.) This correction is smaller by ${\mathcal{O}}(1/\ln
l_m)$ than the leading term, so that we should have
\[
\frac{\beta l_m}{\ln^2l_m}=\frac{1}{2 w^2}+{\mathcal{O}}\left(\frac{1}{\ln
l_m}\right)
\]
and a plot of $\beta l_m/\ln^2l_m$ against $1/\ln l_m$ should show a
linear approach to the limiting value $1/(2 w^2)$.  The results
shown in the inset of Fig.~\ref{fig:beta_scale_lm} are compatible with
this. The plot still shows some curvature, however, indicating that
even our largest values of $l_m=5000$ are not yet large enough for
higher-order corrections to be negligible. Accordingly, a linear
regression gives an intercept of $\simeq 1.81$, rather below the
theoretical value $1/2 w^2=2$ for the chosen $w=0.5$, while a
cubic regression gives a much closer result of $\simeq 1.97$. It is
worth stressing that the above results support the assumption $\beta
l_m\to\infty$ for $l_m\to\infty$ which we made in our theory: the good fit
to the theoretical asymptotics implies that $\beta l_m \sim \ln^2 l_m$
for large $l_m$, which indeed diverges. Even for smaller $l_m$ the
trend for $\beta l_m$ to grow with $l_m$ is noticeable; \eg\ for the
scenario shown in Fig.~\ref{fig:beta_scale_lm}, $\beta l_m$ increases
from $\approx42$ at $l_m=50$ to $\approx431$ at $l_m=5000$.
\bfig
\begin{picture}(0,0)%
\includegraphics{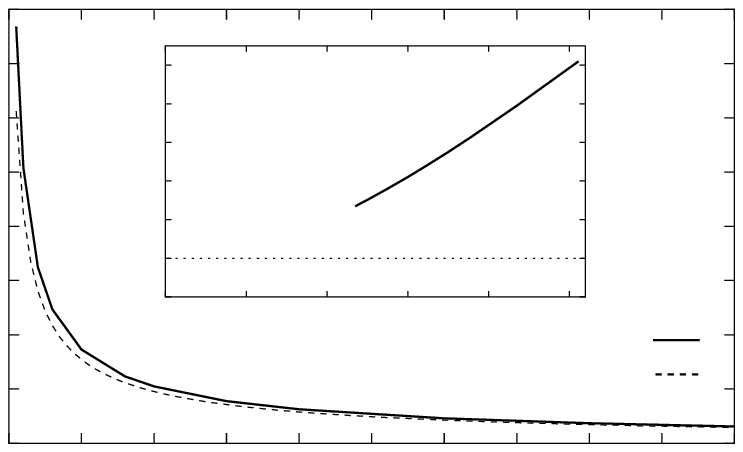}%
\end{picture}%
\setlength{\unitlength}{2565sp}%
\begingroup\makeatletter\ifx\SetFigFont\undefined
% extract first six characters in \fmtname
\def\x#1#2#3#4#5#6#7\relax{\def\x{#1#2#3#4#5#6}}%
\expandafter\x\fmtname xxxxxx\relax \def\y{splain}%
\ifx\x\y   % LaTeX or SliTeX?
\gdef\SetFigFont#1#2#3{%
  \ifnum #1<17\tiny\else \ifnum #1<20\small\else
  \ifnum #1<24\normalsize\else \ifnum #1<29\large\else
  \ifnum #1<34\Large\else \ifnum #1<41\LARGE\else
     \huge\fi\fi\fi\fi\fi\fi
  \csname #3\endcsname}%
\else
\gdef\SetFigFont#1#2#3{\begingroup
  \count@#1\relax \ifnum 25<\count@\count@25\fi
  \def\x{\endgroup\@setsize\SetFigFont{#2pt}}%
  \expandafter\x
    \csname \romannumeral\the\count@ pt\expandafter\endcsname
    \csname @\romannumeral\the\count@ pt\endcsname
  \csname #3\endcsname}%
\fi
\fi\endgroup
\begin{picture}(5454,3802)(659,-3336)
\put(3226,-3286){\makebox(0,0)[lb]{\smash{\SetFigFont{8}{9.6}{rm}{\color[rgb]{0,0,0}$l_m$}%
}}}
\put(1126,-286){\makebox(0,0)[lb]{\smash{\SetFigFont{8}{9.6}{rm}{\color[rgb]{0,0,0}$\frac{\beta l_m}{\ln^2l_m}$}%
}}}
\put(4840,-2359){\makebox(0,0)[lb]{\smash{\SetFigFont{8}{9.6}{rm}{\color[rgb]{0,0,0}$\frac{\ln^2l_m}{2l_m w^2}$}%
}}}
\put(5055,-2114){\makebox(0,0)[lb]{\smash{\SetFigFont{8}{9.6}{rm}{\color[rgb]{0,0,0}$\beta$}%
}}}
\put(3451,-2101){\makebox(0,0)[lb]{\smash{\SetFigFont{8}{9.6}{rm}{\color[rgb]{0,0,0}$1/\ln l_m$}%
}}}
\put(659,-2875){\makebox(0,0)[rb]{\smash{\SetFigFont{8}{9.6}{rm}{\color[rgb]{0,0,0}0}%
}}}
\put(659,-2474){\makebox(0,0)[rb]{\smash{\SetFigFont{8}{9.6}{rm}{\color[rgb]{0,0,0}0.1}%
}}}
\put(659,-2074){\makebox(0,0)[rb]{\smash{\SetFigFont{8}{9.6}{rm}{\color[rgb]{0,0,0}0.2}%
}}}
\put(659,-1673){\makebox(0,0)[rb]{\smash{\SetFigFont{8}{9.6}{rm}{\color[rgb]{0,0,0}0.3}%
}}}
\put(659,-1273){\makebox(0,0)[rb]{\smash{\SetFigFont{8}{9.6}{rm}{\color[rgb]{0,0,0}0.4}%
}}}
\put(659,-872){\makebox(0,0)[rb]{\smash{\SetFigFont{8}{9.6}{rm}{\color[rgb]{0,0,0}0.5}%
}}}
\put(659,-472){\makebox(0,0)[rb]{\smash{\SetFigFont{8}{9.6}{rm}{\color[rgb]{0,0,0}0.6}%
}}}
\put(659,-72){\makebox(0,0)[rb]{\smash{\SetFigFont{8}{9.6}{rm}{\color[rgb]{0,0,0}0.7}%
}}}
\put(659,329){\makebox(0,0)[rb]{\smash{\SetFigFont{8}{9.6}{rm}{\color[rgb]{0,0,0}0.8}%
}}}
\put(1269,-2999){\makebox(0,0)[b]{\smash{\SetFigFont{8}{9.6}{rm}{\color[rgb]{0,0,0}500}%
}}}
\put(1805,-2999){\makebox(0,0)[b]{\smash{\SetFigFont{8}{9.6}{rm}{\color[rgb]{0,0,0}1000}%
}}}
\put(2340,-2999){\makebox(0,0)[b]{\smash{\SetFigFont{8}{9.6}{rm}{\color[rgb]{0,0,0}1500}%
}}}
\put(2876,-2999){\makebox(0,0)[b]{\smash{\SetFigFont{8}{9.6}{rm}{\color[rgb]{0,0,0}2000}%
}}}
\put(3412,-2999){\makebox(0,0)[b]{\smash{\SetFigFont{8}{9.6}{rm}{\color[rgb]{0,0,0}2500}%
}}}
\put(3948,-2999){\makebox(0,0)[b]{\smash{\SetFigFont{8}{9.6}{rm}{\color[rgb]{0,0,0}3000}%
}}}
\put(4484,-2999){\makebox(0,0)[b]{\smash{\SetFigFont{8}{9.6}{rm}{\color[rgb]{0,0,0}3500}%
}}}
\put(5019,-2999){\makebox(0,0)[b]{\smash{\SetFigFont{8}{9.6}{rm}{\color[rgb]{0,0,0}4000}%
}}}
\put(5555,-2999){\makebox(0,0)[b]{\smash{\SetFigFont{8}{9.6}{rm}{\color[rgb]{0,0,0}4500}%
}}}
\put(6091,-2999){\makebox(0,0)[b]{\smash{\SetFigFont{8}{9.6}{rm}{\color[rgb]{0,0,0}5000}%
}}}
\put(733,-2999){\makebox(0,0)[b]{\smash{\SetFigFont{8}{9.6}{rm}{\color[rgb]{0,0,0}0}%
}}}
\put(1846,-1769){\makebox(0,0)[rb]{\smash{\SetFigFont{8}{9.6}{rm}{\color[rgb]{0,0,0}1.9}%
}}}
\put(1846,-1198){\makebox(0,0)[rb]{\smash{\SetFigFont{8}{9.6}{rm}{\color[rgb]{0,0,0}2.1}%
}}}
\put(1846,-627){\makebox(0,0)[rb]{\smash{\SetFigFont{8}{9.6}{rm}{\color[rgb]{0,0,0}2.3}%
}}}
\put(1846,-57){\makebox(0,0)[rb]{\smash{\SetFigFont{8}{9.6}{rm}{\color[rgb]{0,0,0}2.5}%
}}}
\put(1889,-1891){\makebox(0,0)[b]{\smash{\SetFigFont{8}{9.6}{rm}{\color[rgb]{0,0,0}0}%
}}}
\put(2486,-1891){\makebox(0,0)[b]{\smash{\SetFigFont{8}{9.6}{rm}{\color[rgb]{0,0,0}0.05}%
}}}
\put(3083,-1891){\makebox(0,0)[b]{\smash{\SetFigFont{8}{9.6}{rm}{\color[rgb]{0,0,0}0.1}%
}}}
\put(3679,-1891){\makebox(0,0)[b]{\smash{\SetFigFont{8}{9.6}{rm}{\color[rgb]{0,0,0}0.15}%
}}}
\put(4275,-1891){\makebox(0,0)[b]{\smash{\SetFigFont{8}{9.6}{rm}{\color[rgb]{0,0,0}0.2}%
}}}
\put(4872,-1891){\makebox(0,0)[b]{\smash{\SetFigFont{8}{9.6}{rm}{\color[rgb]{0,0,0}0.25}%
}}}
\end{picture}
\caption{Plot of $\beta$ (solid) and the theoretically predicted
asymptotic scaling $\beta=(\ln^2l_m)/(2 w^2)$ (dashed) against $l_m$ at
$w=0.5$. The inset provides a more stringent test of the theory
by plotting $\beta l_m/\ln^2l_m$ versus $1/\ln l_m$. Theory predicts
that this should approach $1/2 w^2=2$ (dotted line) linearly; the
results plotted are broadly compatible with this but still show some
curvature, indicating that our values of $l_m$ are still too small to
have reached the asymptotic regime where higher-order corrections can
be neglected.}
\label{fig:beta_scale_lm}
\efig

Finally, our theory also predicts how the parameter $\rho_2\N$, giving the
overall degree of orientational ordering in the nematic shadow phase,
scales with $l_m$. The result is that $\rho_2\N$ should be
proportional to $\beta^{1/2}$ for large $l_m$, so that a plot of these
two quantities against each other as $l_m$ is varied should give a
straight line through the origin, with slope $\kappa^{-1/2}$ from
Eq.~\eqref{eq:pressure_for_beta}. Fig.~\ref{fig:rho2_beta_new} is
consistent with this, but shows the expected deviations in the
pre-asymptotic regime where $\rho_2\N$ and $\beta$ are not yet small.
\bfig
\begin{picture}(0,0)%
\includegraphics{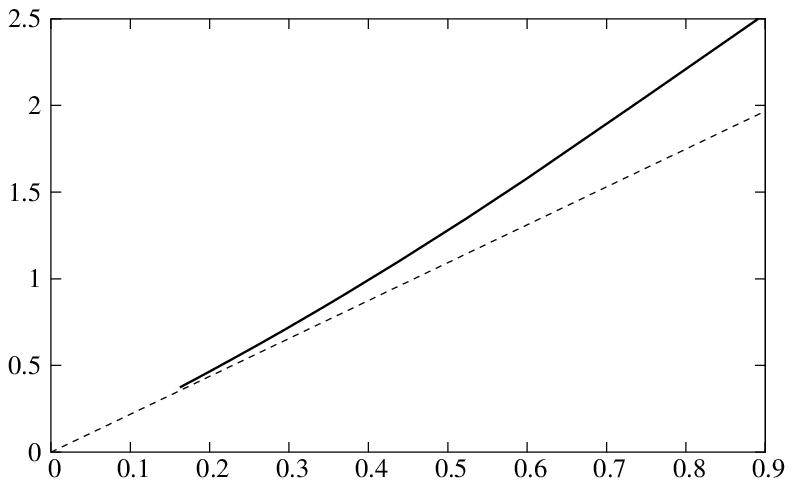}%
\end{picture}%
\setlength{\unitlength}{2565sp}%
\begingroup\makeatletter\ifx\SetFigFont\undefined
% extract first six characters in \fmtname
\def\x#1#2#3#4#5#6#7\relax{\def\x{#1#2#3#4#5#6}}%
\expandafter\x\fmtname xxxxxx\relax \def\y{splain}%
\ifx\x\y   % LaTeX or SliTeX?
\gdef\SetFigFont#1#2#3{%
  \ifnum #1<17\tiny\else \ifnum #1<20\small\else
  \ifnum #1<24\normalsize\else \ifnum #1<29\large\else
  \ifnum #1<34\Large\else \ifnum #1<41\LARGE\else
     \huge\fi\fi\fi\fi\fi\fi
  \csname #3\endcsname}%
\else
\gdef\SetFigFont#1#2#3{\begingroup
  \count@#1\relax \ifnum 25<\count@\count@25\fi
  \def\x{\endgroup\@setsize\SetFigFont{#2pt}}%
  \expandafter\x
    \csname \romannumeral\the\count@ pt\expandafter\endcsname
    \csname @\romannumeral\the\count@ pt\endcsname
  \csname #3\endcsname}%
\fi
\fi\endgroup
\begin{picture}(5800,3769)(1338,-4205)
\put(1390,-1529){\makebox(0,0)[lb]{\smash{\SetFigFont{8}{9.6}{rm}{\color[rgb]{0,0,0}$\rho_2\N$}%
}}}
\put(4245,-4142){\makebox(0,0)[lb]{\smash{\SetFigFont{8}{9.6}{rm}{\color[rgb]{0,0,0}$\sqrt\beta$}%
}}}
\end{picture}
\caption{Plot of $\rho_2\N$ against $\sqrt{\beta}$, obtained by
varying $l_m$ at fixed $w=0.5$ (solid), together with the
asymptotic scaling $\rho_2\N = (\beta/\kappa)^{1/2}$ predicted from
Eq.~\eqref{eq:pressure_for_beta} (dashed). As predicted by theory, both 
$\beta$ and $\rho_2\N$ tend towards zero as $l_m$ increases. In the
same limit they also 
approach the predicted linear asymptote, while
for smaller $l_m$ and therefore larger $\rho_2\N$
and $\beta$ deviations from linearity occur.}
\label{fig:rho2_beta_new}
\efig

\section{Conclusion}
\label{sec:conclusion}

We have analysed the effects of length polydispersity on the phase
behaviour of hard rods, within the $\Ptwo$ Onsager model which is a
simplified version of the full Onsager theory. We focussed in
particular on rod length distributions with fat tails that decay less
than exponentially, using log-normal distributions as our primary
example. Our work was originally motivated by the question of whether,
even for such unimodal length distributions, the presence of
relatively many long rods could cause I-N-N phase separation to occur.
We do indeed find an I-N-N coexistence region in the phase diagram for
a log-normal length distribution with a finite cutoff. Unusually, this
region is located inside the I-N coexistence region, \ie\ not
connected to an N-N region, and very narrow in density, separating two
different regimes of isotropic-nematic coexistence. At densities below
the three-phase region, the nematic phase is significantly enriched in
long rods compared to the isotropic: its density distribution exhibits
a second peak for large rod lengths, in addition to the one at lengths
of order unity (the average rod length in the system). Because the
overall number of long rods in the system is small, however, the
nematic phase only occupies a very small part of the system volume in
this regime. As the density increases towards the three-phase region,
the second peak in the nematic length distribution moves towards the
main peak; beyond the three-phase region, the coexisting nematic only
exhibits a single peak, and the phase behaviour resembles that found
for systems without fat-tailed length distributions. Moving towards
larger polydispersities in the phase diagram, the three-phase region
eventually closes off and is replaced by a crossover, within the I-N
coexistence region, between the two types of nematic phase just
described.  Coming from small polydispersity $\sigma$, on the other
hand, the three-phase region begins at a threshold value of $\sigma$
where the isotropic cloud curve shows a kink and the properties of the
coexisting nematic shadow phase change discontinuously.

In the theoretical part of the paper, we gave an analytical treatment
of the onset of nematic order, \ie\ of the isotropic cloud point, in
the limit of large length cutoffs $l_m$. The key quantities for the theory
were the parameter $\beta$, which determines the exponential enrichment
of the nematic phase in the longer rods, and the average ordering
$\rho_2\N$ of the nematic phase. We were able to show that both of
these quantities must converge to zero for $l_m\to\infty$, and
based on our numerical results supplemented this with the assumption
that $\beta l_m\to\infty$.
On this basis we could construct a self-consistent theory for the
limit of large $l_m$, by treating separately the behaviour of short
rods (with lengths of order one) and long rods of order the cutoff
length. Surprisingly, the theory predicts that the density (and
rescaled rod volume fraction) $\rho_0$ of the isotropic phase, and the
density $\rho_0\N$ and rescaled rod volume fraction $\rho_1\N$ of the
nematic shadow phase, all converge to the same limiting curve when
plotted against polydispersity $\sigma$. This curve can be calculated
explicitly and depends on no other details of the rod length
distribution except for the presence of the fat tail. The theory in
fact makes a stronger statement about the similarity of the isotropic
cloud and nematic shadow phases: for any rod length of order unity,
the density distributions in the two phases become equal for
$l_m\to\infty$, and the degree of orientational order of such rods in
the nematic shadow tends to zero. This does not, however, imply that
the isotropic and nematic phases coincide completely, or that the
nematic lacks orientational order: when we consider the longest rods
in the system, of the order of the cutoff length, we find that the
nematic shadow contains a much larger number of them, which also
exhibit strong nematic order.
Our theory also allowed us ascertain the rate of convergence of all
quantities to their limit values, to leading order as $l_m\to\infty$,
and the predictions were in good agreement with numerical results.

An important question that our current study cannot answer directly is
how the above results for the $\Ptwo$ Onsager model will carry over to
the full Onsager theory. The comparison of our earlier results for
bidisperse distributions~\cite{SpeSol_p2} were encouraging in this
respect, showing qualitative agreement except for the presence of an
N-N critical point in the $\Ptwo$ Onsager model which in the Onsager
theory is replaced by an unbounded N-N coexistence region opening
towards large densities. A direct attack on the full Onsager problem,
with the required self-consistent determination of all orientation
distributions, is of course a formidable numerical problem.
Nevertheless, we have recently obtained preliminary results which
suggest that the isotropic cloud curve for Onsager theory with a
log-normal length distribution indeed features a kink, as it does for
the $\Ptwo$ Onsager model studied above; this shows indirectly that
the full Onsager theory must also exhibit an I-N-N coexistence region
as found above. The overall topology of the phase diagram remains an
open question, however; in particular, one would like to know whether
the I-N-N coexistence is again confined to a narrow interval or
whether it extends over a broader range and is bordered by an N-N
phase separation at higher densities.

{\bf Acknowledgements:} PS is grateful for financial support through
EPSRC grant GR/R52121/01.

%%%%%%%%%%%%%%%%%%%%APPENDIX%%%%%%%%%%%%%%%%%%%%%%%%%%%%%%%%%
\begin{appendix}
\section{The case $\beta\rightarrow 0$, $\rho_2\N $ finite}
\label{app:rho2N=0}

In our theory for the isotropic cloud point we showed that in the
limit of large cutoff $l_m\to\infty$, the parameter $\beta$ tends
to zero while $\beta l_m\to\infty$. The moment density $\rho_2\N$, on
the other hand, which describes the orientational ordering of the
nematic shadow phase, could in principle either also tend to zero, or
converge to a nonzero constant in the limit. In this appendix we show
that the second case can be excluded because it does not lead to a
consistent solution of the cloud point equations.

The assumption that $\beta\to 0$ again allows us to break all
integrals over $l$ into two intervals, $l=0\ldots\tilde{l}$ and
$l=\tilde{l}\ldots l_m$ with $1\ll\tilde{l}\ll l_m$. All statements
made in the main text regarding the long rod contributions $I_0$,
$I_1$ and $I_2$ to the nematic shadow phase moments $\rho_0\N$,
$\rho_1\N$ and $\rho_2\N$ also remain valid, seeing as the only
property of $\rho_2\N$ that entered was that $\rho_2\N l_m$ diverges
as $l_m\to\infty$; if $\rho_2\N$ itself has a nonzero limit value then
this is certainly the case. We know therefore that $I_0$ is smaller by
a factor $1/l_m$ than $I_1$ and can be neglected, and that $I_2=I_1$
to leading order 
since as a consequence of $\rho_2\N l_m\to\infty$ the longest rods are
strongly ordered.

The treatment of the short rod parts is actually easier than in the
case of $\rho_2\N\to 0$: a finite limit value of $\rho_2\N$ ensures
that these integrals have nonvanishing differences from the
corresponding values in the isotropic phase, even for
$l_m\to\infty$. Any corrections from the small nonzero value of
$\beta$ are negligible by comparison, so that we can set $\beta=0$
directly and write Eqs.\eqqq{eq:rho0N}{eq:rho1N}{eq:rho2N} as 
%\[
\begin{eqnarray}%{lcl}
\rho_0\N &=&\rho_0\In_0 \nonumber\\
\rho_1\N &=&\rho_0\left(\In_1+\tI_1\right) \nonumber\\
\rho_2\N &=&\rho_0\left(\In_2+\tI_1\right) \label{rho2N_app}
\end{eqnarray}
%\]
where $\tI_1=I_1/\rho_0$ and 
%\[
\begin{eqnarray*}%{lcl}
\In_{0}&=&\totint\normparent(l)e^{f(\theta)l}\\
\In_{1}&=&\totint l\normparent(l)e^{f(\theta)l}\\
\In_{2}&=&\totint l\normparent(l)\Ptwo e^{f(\theta)l}
\end{eqnarray*}
%\]
%
(In principle, the short rod integrals run over $l=0\ldots\tilde{l}$,
but because the integrals as written are convergent, one can take the
upper limit to $\infty$ without error at the level of our approximation.)
We can determine the value of $\tI_1$ from Eq.~\eqref{eq:beta} for $\beta$,
which using $\beta\to 0$ becomes 
\[
0=-c_1\left(\rho_1\N -\rho_0\right)+c_2\rho_2\N 
\]
and solving for $\tI_1$ one finds
\beq
\tI_1=\frac{c_1-c_1\In_1+c_2\In_2}{c_1-c_2}
\label{I1tilde}
\eeq
Inserting this into the pressure equality~\eqref{eq:osmotic_pressure},
which with a common factor of $\rho_0$ removed reads
\[
1+\frac{c_1}{2}\rho_0=\In_0+\frac{c_1}{2}\rho_0(\In_1+\tI_1)^2
-\frac{c_2}{2}\rho_0(\In_2+\tI_1)^2
\]
we can solve explicitly to obtain the cloud point density $\rho_0$ as
\beq
\label{eq:rho0_app}
\rho_0 = \frac{c_1-c_2}{c_1 c_2} \frac{2(1-\In_0)}{1-(\In_1-\In_2)^2}
\eeq
where the r.h.s.\ is now a function only of
$\rho_2\N$, via the short rod integrals $\In_0$, $\In_1$ and $\In_2$.
Finally, substituting the results~\eqref{I1tilde} for $\tI_1$
and~\eqref{eq:rho0_app} for $\rho_0$ into Eq.~\eqref{rho2N_app} for
$\rho_2\N $, we obtain after a little algebra the
self-consistency equation
\beq\label{eq:fix_point}
\rho_2\N =F(\rho_2\N )
\eeq
where 
\[
F(\rho_2\N )=\frac{2(1-{\In_0})}{c_2({\In_1}-{\In_2}+1)}
\]
It can be shown that this function passes through the origin with a
slope of one, but then curves downwards and approaches a constant for
large $\rho_2\N$. It follows that
\[
%\frac{d F}{d\rho_2\N }\leq 1
F(\rho_2\N )<\rho_2\N 
\]
%@@FOR THESIS ONLY:\\
%Need proof of this for thesis@@\\
for $\rho_2\N > 0$, so that
%\bfig
%\input{fix_point}
%\caption{Solution of the fixed point equation~\eqref{eq:fix_point} for a
%log-normal parent, with $\sigma=0.5$. It is easy to show that
%$\rho_2\N =0$ is the only solution to such an equation for any parent
%distribution.}
%\label{fig:fix_point}
%\efig
%%
the only solution is $\rho_2\N=0$, in contradiction to our assumption
that $\rho_2\N$ had a nonzero limit value for $l_m\to\infty$. This
assumption must therefore be abandoned, and we conclude that
$\rho_2\N$ must tend to zero for $l_m\to\infty$. From our arguments it
is clear that this conclusion holds for any fat-tailed parent
distribution.

We mention finally that, even though the above treatment based on the
assumption of nonzero $\rho_2\N$ is not fully self-consistent, it
actually recovers the correct limiting value of the cloud point
density. As pointed out above, Eq.~\eqref{eq:fix_point} has
$\rho_2\N=0$ as its only solution. Now for small $\rho_2\N$ one can
expand  $\In_0=1-c_2 \rho_2\N$,
$\In_1=1-c_2\rho_2\N\langle l^2\rangle$ and $\In_2=(c_2/5)\rho_2\N
\langle l^2\rangle$; inserting these results into
Eq.~\eqref{eq:rho0_app} one finds $\rho_0=1/(4\langle l^2\rangle)$ in
the limit $\rho_2\N\to 0$, in agreement with Eq.~\eqref{eq:CPC_rho_0}.

\section{The long rod integrals}
\label{app:long_rods_integrals}

In this appendix we discuss how to evaluate the general long rod integral
\[
I= \rho_0 \int_{\tilde l}^{l_m} dl\ P(l)l^n e^{\beta
l}\angint e^{lf(\theta)} 
\]
with $P(l)\equiv \normparent(l)$ for brevity. For $n=0$ and $1$ this
reduces to $I_0$ and $I_1$, the long rod contributions to the moments
$\rho_0\N$ and $\rho_1\N$ of the nematic shadow phase. As explained
before Eq.~\eqref{eq:nematic_moments_sno}, since the integral will
turn out to be dominated by the long rods with $l\approx l_m$, we can
use the fact that these rods are strongly ordered ($\rho_2\N l\gg 1$)
to perform the angular integral and get
\beq
I = \frac{\rho_0}{3c_2\rho_2\N}\II, \qquad
\II = \int_{\tilde l}^{l_m}dl\ P(l) l^{n-1} e^{\beta l}
\label{eq:II_app}
\eeq

Our task now is to evaluate $\II$. If this really is dominated by
$l\approx l_m$ due to the presence of the exponential factor
$\exp(\beta l)$, we should be able to replace the rest of the
integrand, $f(l)=P(l) l^{n-1}$ by its value at $l=l_m$. To check the
quality of this approximation, we include the first term in a Taylor
expansion,
\beq\label{eq:f(xi)_expansion}
f(l) = f(l_m)+(l-l_m) f^\prime(l_m) + \ldots
\eeq
which when inserted into Eq.~\eqref{eq:II_app} gives
\begin{equation}\label{eq:expansion_integral}
\II =
\left[f(l_m)-\frac{f^\prime(l_m)}{\beta}\right]\frac{e^{\beta l_m}}{\beta}
%\frac{f(l_m)}{\beta}e^{\beta
%l_m}+f^\prime(l_m)\frac{e^{\beta l_m}}{\beta
%}l_m\left[1-\frac{1}{\beta l_m}-1\right] \\\nonumber
%&=&
\end{equation}
If the leading term is a good approximation, the correction term
should be small, \ie\ $-f^\prime(l_m)/[\beta f(l_m)]\to 0$ for
$l_m\to\infty$. Whether this is true depends on the $l_m$-dependence
of $\beta$. From Eq.~\eqref{eq:beta_lm}, this is given implicitly by
$\II/\beta = A$, with $A$ a constant and $\II$ evaluated at $n=1$,
\ie\ for $f(l)=P(l)$. Keeping the leading term of $\II$ only,
we get
\beq\label{eq:F(beta l_m)}
\frac{e^{\beta l_m}}{(\beta l_m)^2} = \frac{A}{P(l_m) l_m^2}
%\frac{A}{f(l_m) l_m^2}=\frac{e^{\beta l_m}}{\beta l_m}
\eeq
Using the inverse of the function $G(x)=e^x/x^2$ on the l.h.s.\ this
can be formally solved to give
\[
\beta=\frac{1}{l_m}G^{-1}\left(\frac{A}{P(l_m)l_m^2}\right)
\]
Now from the definition of $G(x)$ we have
\[
G(x)=y \quad \Rightarrow \quad x=\ln y+2\ln x
\]
and therefore by recursion
\[
x=\ln y+2\ln(\ln y+ 2\ln(\ln y + 2\ln\ldots))
\]
Asymptotically the first term is dominant, so that $x=G^{-1}(y)=\ln
y$ and the asymptotic behaviour of $\beta$ becomes
\beq
\label{eq:beta_app}
\beta=\frac{1}{l_m}\ln\left(\frac{A}{P(l_m)l_m^2}\right)
\eeq
The condition $-f'(l_m)/[\beta f(l_m)]\to 0$ for our leading
order-approximation for $\II$ to be correct is thus
\beq\label{eq:function_to_0}
-\frac{l_m f^\prime(l_m)}{f(l_m)\{\ln A - \ln[P(l_m)l_m^2]\}} 
\rightarrow 0
\eeq
If we now restrict our attention to
parent distributions with finite mean, \ie\ $\lint P(l) l<\infty$, then
$P(l_m)l_m^{2}\to 0$ for $l_m\to\infty$ and
the constant term $\ln A$ becomes negligible. We therefore need
\beq
\label{eq:ratio}
\frac{l_m (\ln f)^\prime(l_m)}{\ln[P(l_m)l_m^2]} =
\frac{l_m [(\ln P)^\prime(l_m)+(n-1)/l_m]}{\ln[P(l_m)l_m^2]} 
\rightarrow 0
\eeq
For a log-normal parent distribution, $\ln P(l) = -(\ln^2 l)/ 
(2 w^2)$ for large $l$ and so $(\ln P)'(l) = - (\ln l)/(w^2 l)$,
and the condition becomes
\[
\frac{l_m [(\ln l_m)/l_m + \order(l_m^{-1})]}
{(\ln^2 l_m)^2/2 + \order(\ln l_m)} \simeq \frac{2}{\ln l_m} \to 0
\]
which is obviously fulfilled for $l_m\to\infty$. For power-law
parents, $\ln P(l) \sim \ln l$, one also finds that the
ratio~\eqref{eq:ratio} tends to zero logarithmically. In both these
cases, one is therefore justified in only keeping the leading term in
our Taylor expansion around $l=l_m$, giving
\[
\II = \frac{f(l_m)e^{\beta l_m}}{\beta} =
\frac{P(l_m)l_m^{n-1}e^{\beta l_m}}{\beta}
\]
This proves in particular the statement made in the main text that the
effect of the weight function $l^n$ only comes in through the factor
$l_m^n$. It is also worth stressing that Eq.~\eqref{eq:beta_app}
implies that $\beta l_m \simeq -\ln[P(l_m)l_m^2] \to\infty$,
consistent with our assumption of such a divergence. This result also
justifies a posteriori the intuition that the integrals for $I$ and
$\II$ are dominated by the longest rods: the exponential integral
giving the leading order approximation has significant contributions
only where $l_m-l$ is of order $1/\beta$, so that the fractional
deviation of the relevant $l$-values from $l_m$, $(l_m-l)/l_m \sim
1/(\beta l_m)$ tends to zero for $l_m\to\infty$. 

In the more general case of a less than exponentially decaying parent,
say $P(l)= e^{-h(l)}$ and $h(l)\sim l^\alpha$ with $0<\alpha<1$, we
have $(\ln P)'(l) \sim l^{\alpha-1}$ and the
condition~\eqref{eq:ratio} becomes
\[
\frac{l_m [\alpha
l_m^{\alpha-1} + \order(l_m^{-1})]}{l_m^{\alpha} + \order(\ln l_m)}
\simeq \alpha\neq 0
\]
Here one therefore needs to go beyond the leading order approximation
for $\II$. To do so, one can write $f(l)=\exp[-h(l)+(n-1)\ln l]$ and keep
in Eq.~\eqref{eq:II_app} linear terms in the Taylor expansion of the
exponent around $l=l_m$. This results in
\begin{eqnarray}
\II &=& \int_{\tilde l}^{l_m}dl\ e^{-h(l_m)+(n-1)\ln l_m+\beta l_m +
(l-l_m)[\beta-h'(l_m)+(n-1)/l_m]}
\nonumber\\
&=& \frac{l_m^{n-1}e^{-h(l_m)+\beta l_m}}{\beta-h'(l_m)+(n-1)/l_m}
\label{eq:II_refined_app}
\end{eqnarray}
For $n=1$ this gives Eq.~\eqref{eq:II_refined} in the main text. (The
term $(n-1)/l_m$ in the denominator can be neglected even for $n\neq
1$ since, having dealt with the power-law distributions case $h(l)\sim
\ln l$ above, we can here assume that $h(l)$ diverges more strongly
than $\ln l$.) From this one can again deduce the scaling of $\beta$,
using the condition $\II/\beta = A$ 
\[
\frac{e^{-h(l_m)+\beta l_m}}{\beta[\beta-h'(l_m)]} = A
\]
To leading order this gives Eq.~\eqref{eq:beta_scaling_general},
$\beta=h(l_m)/l_m$; this can be shown by writing
$\beta=h(l_m)/l_m+\delta$, substituting into the above condition and
verifying that the assumption that $\delta$ is a small correction is
self-consistent. Once the scaling of $\beta$ is established, one can
then show that higher-order corrections to the first-order
approximation~\eqref{eq:II_refined_app} are indeed vanishingly small
as $l_m\to\infty$; we omit the details here. \\

\end{appendix}

\bibliographystyle{unsrt}
\bibliography{references}

\begin{thebibliography}{10}
\bibitem{norm_parent}
In the following we will use the same convention as we used in the numerical
  solution of the phase coexistence equations, \ie, we will assume that the
  functional form of $\normparent(l)$ is fixed as $l_m$ is varied. In this way
  the parent will be strictly normalized only in the limit for
  $\l_m\rightarrow\infty$, and the average length will be strictly equal to 1
  in the same limit. However, as we mentioned in Sec.~\ref{sec:numerical_res},
  the corrections to the strict normalization are small for sufficiently large
  cutoff. In the following, as we are interested in the large cutoff limit, we
  will therefore always assume $\rho_1=\rho_0$ for the parent distribution.
\end{thebibliography}
\vfill\eject
\end{document}